\documentclass[
preprint,
 amsmath,amssymb,
 aps,
]{revtex4-2}

\usepackage[T1]{fontenc}
\usepackage[utf8]{inputenc}
\usepackage{graphicx}
\usepackage{dcolumn}
\usepackage{bm}
\usepackage{mathtools}
\usepackage{amsmath}
\usepackage{ragged2e}
\usepackage{longtable}
\usepackage[usenames,dvipsnames]{color}
\usepackage{float}
\usepackage{siunitx}
\usepackage{accents}
\usepackage[dvipsnames]{xcolor}
\usepackage{braket}

\begin{document}


\title{Suppression of Spectral Gap and Flat Bands on a Cuprate Superconductor Side-Surface}

\author{Gabriele Domaine\textsuperscript{1,2}}


\author{Mihir Date\textsuperscript{1,3}}
\author{Sydney K. Y. Dufresne\textsuperscript{1}}
\author{Natalie Lehmann\textsuperscript{1}}
\author{Daiyu Geng\textsuperscript{1}}
\author{Tohru Kurosawa\textsuperscript{4}}
\author{Amit Kumar\textsuperscript{1}}
\author{Jiaju Wang\textsuperscript{1}}
\author{Tianlun Yu\textsuperscript{1}}
\author{Chien-Ching Chang\textsuperscript{1}}
\author{Swosti P. Sarangi\textsuperscript{1}}
\author{Ding Pei\textsuperscript{5}}
\author{Yiran Liu\textsuperscript{2}}
\author{Julia K\"{u}spert\textsuperscript{6}}
\author{Shigemi Terakawa\textsuperscript{1,7,8}}
\author{Markel Pardo Almanza\textsuperscript{1}}
\author{Jiabao Yang\textsuperscript{1}}
\author{Izabela Biało\textsuperscript{6}}
\author{Matthew D. Watson\textsuperscript{3}}
\author{Timur K. Kim\textsuperscript{3}}
\author{Stephen M. Hayden\textsuperscript{9}}
\author{Kritika Singh\textsuperscript{10}}
\author{Banabir Pal\textsuperscript{1}}
\author{Matteo Minola\textsuperscript{2}}
\author{Johan Chang\textsuperscript{6}}
\author{Naoki Momono\textsuperscript{11}}
\author{Migaku Oda\textsuperscript{4}}
\author{Stuart S. P. Parkin\textsuperscript{1}}
\author{Andreas P. Schnyder\textsuperscript{2}}
\author{Niels B. M. Schröter\textsuperscript{1,12, 13}}
\email{niels.schroeter@mpi-halle.mpg.de}
\affiliation{\textsuperscript{1}Max Planck  Institut f\"ur  Mikrostrukturphysik, Weinberg 2, 06120 Halle, Germany}

\affiliation{\textsuperscript{2}Max-Planck-Institut für Festkörperforschung, Heisenbergstrasse 1, D-70569 Stuttgart, Germany}

\affiliation{\textsuperscript{3}Diamond Light Source Ltd, Harwell Science and Innovation Campus, Didcot, OX11 0DE, United Kingdom}

\affiliation{\textsuperscript{4}Department of Applied Physics, Hokkaido University, Sapporo 060-8628, Japan}

\affiliation{\textsuperscript{5}Synchrotron SOLEIL, L’Orme des Merisiers, Saint-Aubin, 91190, France}

\affiliation{\textsuperscript{6}Physik-Institut, Universität Zürich, Winterthurerstrasse 190, CH-8057, Zürich, Switzerland}

\affiliation{\textsuperscript{7}Department of Applied Physics, Graduate School of Engineering, The University of Osaka, Osaka
565-0871, Japan}

\affiliation{\textsuperscript{8}Center for Future Innovation, Graduate School of Engineering, The University of Osaka, Osaka
565-0871, Japan}



\affiliation{\textsuperscript{9}H. H. Wills Physics Laboratory, University of Bristol, Bristol BS8 1TL, United Kingdom}

\affiliation{\textsuperscript{10}Institute of Materials Physics, Helmholtz-Zentrum Hereon, Geesthacht, 21502, Germany}

\affiliation{\textsuperscript{11}Faculty of Science and Engineering, Muroran Institute of Technology, Muroran 050-8585, Japan}

\affiliation{\textsuperscript{12}Martin-Luther-Universität Halle-Wittenberg, Institut für Physik, Von-Danckelmann-Platz 3, 06120, Halle (Saale), Germany}

\affiliation{\textsuperscript{13}Halle-Berlin-Regensburg Cluster of Excellence CCE, Halle (Saale), Germany}

\date{\today}

\begin{abstract}

Side surfaces of cuprate superconductors are expected to display a suppressed $d$-wave order parameter and zero-energy topological flat bands with a large density of states, making them susceptible to symmetry broken orders. Yet such surfaces have never been investigated with momentum-resolved, surface-sensitive probes, because high-temperature superconductors rarely cleave along them. Using focused-ion-beam milling to define a controlled breaking point, we expose pristine (110) side surfaces of overdoped La$_{2-x}$Sr$_x$CuO$_4$ ($x=0.22$) suitable for angle-resolved photoemission. We observe the suppression of the superconducting spectral gap within our energy resolution ($\sim 4~\mathrm{meV}$), and surprisingly, the expected zero-energy flat band peak is also suppressed, despite the high topographic quality of the surface. Self-consistent Bogoliubov--de~Gennes calculations show that the measured geometric roughness of the cleaved surface is too weak to eliminate these modes. The calculations further demonstrate that bulk inhomogeneities characteristic of high-temperature superconductors, modelled as moderate Anderson-type disorder, can broaden the flat-band states beyond detectability. Our results provide the first momentum-resolved view of the electronic structure on a cuprate side surface and reveal disorder as the key factor currently preventing appearance of flat bands and their associated correlated orders.

\end{abstract}

\maketitle



\section{\label{sec:level1}Introduction\protect}
\begin{figure}[H]
    \centering
    \includegraphics[width = 1\textwidth]{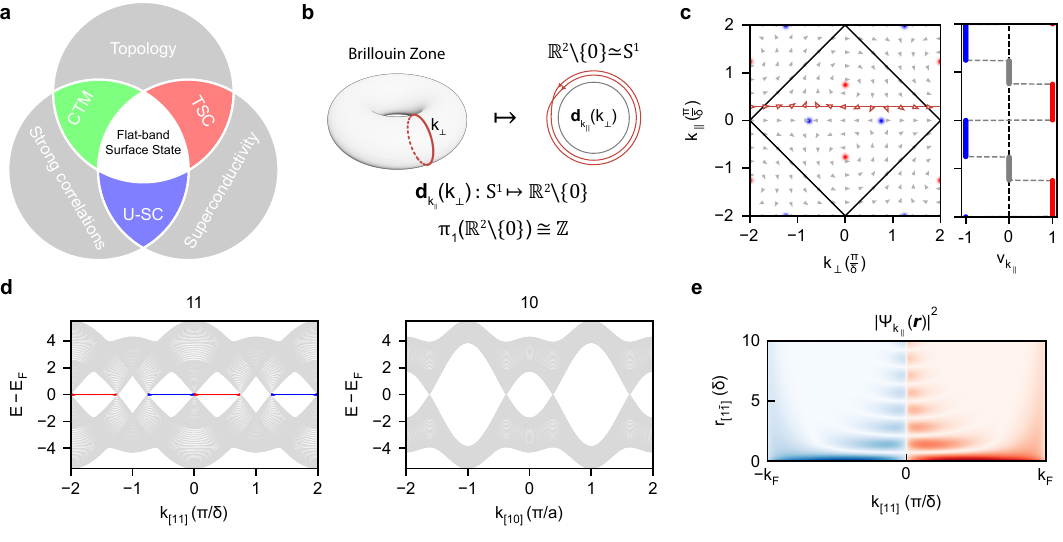}
   \caption{\textbf{Edge states in clean $d_{x^{2}-y^{2}}$-wave superconductors.} \textbf{a}) Illustration of the realization of novel phenomena at the intersection of topology, strong correlations and superconductivity, ranging from topological superconductivity (TSC), unconventional superconductivity (USC) and correlated topological matter (CTM). \textbf{b}) Illustration of the mapping defined by the field $\bm{d}_{k_{\parallel}}$ from one-dimensional submanifolds of the Brillouin zone, which is homeomorphic to a 2-Torus, into the group $\mathbb{R}^{2}\setminus\{0\}$, which is homotopy equivalent to $S^1$. \textbf{c}) Illustration of the vector field $\bm{d}_{\bm{k}}$ for a $d_{x^{2}-y^{2}}$-wave superconductor in the first two Brillouin zones of a square lattice, so that $k_{\perp} = k_{[11]}$ and $k_{\parallel} = k_{[1\bar{1}]}$. The red horizontal line corresponds to a loop in momentum space as in the previous panel. The superconducting nodes of winding number $+1$ and $-1$ are shown in red and blue, respectively. The right inset shows the resulting winding numbers $\nu_{k_{\parallel}}$ computed along $k_{\perp}$ and color coded based on their sign, which also identifies their chirality. \textbf{d}) Band structures for different edge orientations, with the two eigenvalues closest to the center of the spectrum highlighted in red and blue according to their chirality eigenvalues. The figure on the left corresponds to a [11] edge, in the same configuration as in Fig.~\ref{figure_1}c, where the non-trivial values of the winding numbers lead to the formation of zero-energy flat bands connecting the projections of the superconducting nodes of opposite winding numbers. On the other hand, the figure on the right corresponds to a [10] edge, where the winding number is trivial for all $k_{\parallel} = k_{[10]}$. In all the figures we defined $\delta = a\sqrt(2)$, with $a$ the lattice constant of the square lattice. \textbf{e}) Exponential decay of the edge states away from the edge ($r_{[1\bar{1}]} = 0$) and their delocalization in the bulk in correspondence of the projection of the superconducting nodes at $k_{[11]} = -k_{F}$, $0$, $k_{F}$. The analytical expression of the edge state wavefunction is derived in the Supplementary section \ref{sup_sub_sec_1}.}
    \label{figure_1}
\end{figure}

The intersection of superconductivity, topology, and strong correlations underlies many of the most unconventional quantum phases known today~\cite{Regnault_2011,Neupert_2011,Wang_2011,Tang_2011,Liu_2012,Lauchli_2013,Kitaev_2006,Balents_2010,Dzero_2016,Stewart_2017,Keimer_2015,Maeno_2024,Cao_2018,Sato_2017,Elliott_2015,Alicea_2012}. Within this broader landscape, the side surfaces of cuprate $d$-wave superconductors constitute a promising platform for studying the interplay of topology and strong correlations (Fig.~\ref{figure_1}a). Owing to the chiral symmetry of the Bogoliubov-de Gennes Hamiltonian, most of these surfaces host topological flat-band states pinned at the Fermi level~\cite{Hu_1994,Ryu_2002,Sato_2011} (Fig.~\ref{figure_1}b-e). The extensive degeneracy of these states makes them highly susceptible to symmetry-breaking instabilities. This tendency is further amplified by the pair-breaking nature of these surfaces, which suppresses the local superconducting order parameter~\cite{Fogelstr_m_1997,Bruder_1990} and thereby makes it energetically easier for competing orders to emerge. As a result, a variety of symmetry-broken phases have been predicted to emerge, including time-reversal symmetry-breaking states~\cite{Fogelstr_m_1997,Honerkamp_1999, Tomizawa_2008,Nagai_2018}, mixed-parity superconductivity~\cite{Matsubara_2020}, phase-crystal states~\cite{Chakraborty_2022,Wennerdal_2020,H_kansson_2015, Holmvall_2020,Seja_2024}, and magnetic order~\cite{Seja_2024,Potter_2014,Kuboki_2014}. 

Despite this potential and decades of experimental investigations of the cuprates, experimental evidence of order-parameter suppression due to pair-breaking edges remains limited to step-edges ~\cite{Misra_2002,Sharoni_2001}, while direct identifications of topological side-surface states remain limited to zero-bias conductance peaks (ZBCPs) in tunneling measurements~\cite{Covington_1997, Alff_1997, Asulin_2004, Giubileo_2001, Engelhardt_1999, Wang_1999}. However, these signatures have often been subject to ambiguous interpretations. There have been numerous reports of ZBCPs obtained on surfaces that are not expected to support topological surface states~\cite{Lee_1989, Tsai_1989, Kwo_1990, Mandrus_1991, Iguchi_1992,Lesueur_1992}, whereas other studies have failed to observe a ZBCP altogether~\cite{Kohen_2000, Gurvitch_1989, Kane_1994}. Several mechanisms have been proposed to reconcile these disparate findings, including the presence of inter-domain interfaces~\cite{Hu_1998}, disorder-induced bound states~\cite{Sulangi_2018, Kalenkov_2004}, and the coexistence of multiple facets with different orientations~\cite{Fogelstr_m_1997,Sharoni_2001}, but the lack of momentum resolution in tunneling experiments fundamentally limits the ability to distinguish topological flat bands from topologically trivial in-gap states. A surface-sensitive, momentum-resolved probe is therefore needed, as the surface states are confined within a coherence length from the surface and exhibit a characteristic momentum-space distribution not expected for disorder-induced in-gap states. In addition, momentum resolution is essential for accurately identifying and characterizing potential symmetry-broken phases. 



However, such measurements have been hindered by the layered structure of the cuprate superconductors, which prevents clean side surfaces from being obtained by conventional cleavage. Recently, however, a novel sample‐preparation technique for Angle-Resolved Photoemission Spectroscopy (ARPES) was demonstrated on the normal state of Sr\textsubscript{2}RuO\textsubscript{4}~\cite{Hunter_2024}. In this approach, focused‐ion‐beam (FIB) milling is used to mill micro‐notches into the sample in order to enforce a desired cleavage plane perpendicular to the layers. Here, we apply this approach for the first time to a cuprate high-temperature
superconductor. Using FIB milling to define a cleaving plane, we expose pristine (110) side surfaces of overdoped La$_{2-x}$Sr$_x$CuO$_4$ (LSCO) with $x=0.22$ suitable for ARPES, enabling the first momentum-resolved measurements on a cuprate side surface in the superconducting state. So far, ARPES measurements have only investigated the (001) top surface of this compound and found a superconducting gap of 15-25 meV\cite{Zhong_2022}.


Our measurements reveal a collapse of the spectral gap on the (110) surface, but surprisingly, also no peak associated with the predicted flat-band surface states is observed in the raw spectra, despite high instrumental energy and momentum resolution. Fully momentum-resolved, self-consistent Bogoliubov–de Gennes tight-binding calculations indicate that the experimentally measured surface roughness, within the experimental resolution, is insufficient to account for this absence. Moreover, motivated by the characteristic bulk inhomogeneity observed in high-temperature cuprate superconductors~\cite{Chen_2012,Pasupathy_2008,Graser_2007,Tolpygo_1996,Sun_1995}, we introduce moderate Anderson disorder with a magnitude comparable to the superconducting gap into our model and show that such disorder can strongly suppress the flat-band surface states.
Our results show that side-surface ARPES enabled by FIB cleaving provides momentum-resolved access to the electronic structure of previously inaccessible cuprate surfaces, and reveal the importance of both geometric and Anderson-type disorder in the detection of these surface states, laying groundwork for future studies of correlated topological superconductivity.

\begin{figure}[H]
    \centering
    \includegraphics[width = 1\textwidth]{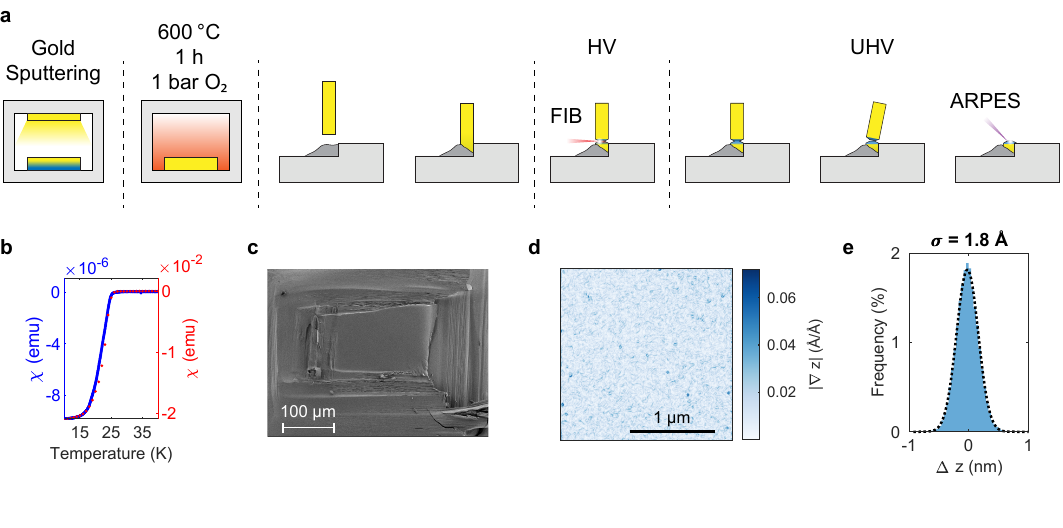}
   \caption{\textbf{Sample preparation} \textbf{a}) Illustration showing the different stages of the sample preparation. \textbf{b}) Temperature dependence of the magnetic susceptibility measured by SQUID magnetometry for a piece of the original rod (red) and of a small part of the milled region of the pillar (blue). Measurements were performed under zero-field-cooled conditions with an applied magnetic field of 5 Oe (red) and 2 Oe (blue), showing no change in the critical temperature during sample preparation. \textbf{c}) Scanning Electron Microscope image of the sample after the mechanical cleave. \textbf{d}) Topography gradient map of the cleaved surface obtained by Atomic Force Microscope and (\textbf{e}) its corresponding height distribution showing a roughness of only $1.8\ \text{\AA}$ (for comparison the in-plane lattice constant of LSCO is $3.78\ \text{\AA}$). The roughness is defined as the root-mean-square deviation of the height profile.}
    \label{figure_2_FIB}
\end{figure}

\section{\label{sec:level3}Results and Discussions}

\subsection{Expected momentum structure of flat band surface states}

The expected momentum structure of the topological side-surface states originates from a topological $\mathbb{Z}$ winding number that the superconducting quasiparticles carry near the nodes of the d-wave order parameter, which is protected by the chiral symmetry of the cuprate superconductors (Fig.~\ref{figure_1}b, see also the Supplementary section \ref{sup_sub_sec_1}). Because the spectrum is gapless, this invariant can only be defined on one-dimensional loops that avoid the nodes. A natural choice is therefore to consider loops indexed by the momentum parallel to the edge, so that one can identify a topological sector --- where the winding is non-zero and requires the existence of zero-energy edge states~\cite{Ryu_2002,Sato_2011} --- and a trivial sector, where the winding vanishes. For a d-wave order parameter, the topological sector is bounded by the projection of two superconducting nodes with opposite winding number (Fig.~\ref{figure_1}c), so that in momentum space the topological states form flat bands that connect the projected superconducting nodes (Fig.~\ref{figure_1}d-e). This leads to a unique alternating momentum structure where the flat bands appear in some but not all of the projected bulk band gaps, becoming the largest for the $[11]$ edge, which motivated us to search for them with ARPES.

\subsection{Surface preparation and characterization}

A schematic of the sample preparation is shown in Fig.~\ref{figure_2_FIB}a. The samples are oriented with Laue and then gold coated and annealed to make a good electrical contact. The samples are then glued, a micro-notch is carved with FIB, and the samples are subsequently cleaved in the ARPES chamber. This process does not affect the bulk superconductivity of the samples, as can be seen from measurements of the magnetic susceptibility before and after the milling process (Fig.~\ref{figure_2_FIB}b). When measuring the magnetic susceptibility after the milling, only a thin slab at the top of the cleaved crystal was used in order to maximize the fraction of milled sample being measured. 

Fig.~\ref{figure_2_FIB}c-e show an Scanning Electron Microscope (SEM) image of a cleaved surface after focused ion beam milling, and the Atomic Force Microscope (AFM) characterization of the surface roughness. The roughness of the cleaved surface, defined as the root-mean-square deviation of the height profile, was determined to be approximately $1.8\ \text{\AA}$, which corresponds to about one half of the in-plane lattice constant of LSCO~\cite{Wang_2019}, or equivalently, to roughly one third of the unit cell dimension along the [110] crystallographic direction. This roughness is comparable to reported values for YBa$_2$Cu$_3$O$_{7-\delta}$ epitaxial thin-films where a uniform ZBCP has been observed with scanning tunneling spectroscopy~\cite{Alff_1997}.

\subsection{Fermi surface characterization}
\begin{figure}[H]
    \centering
    \includegraphics[width = 1\textwidth]{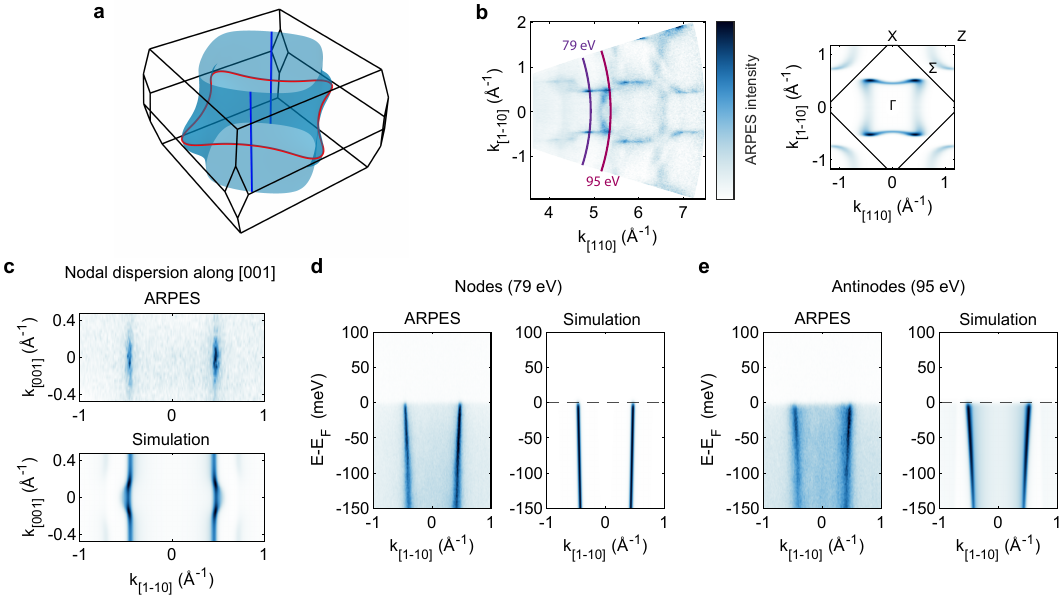}
   \caption{\textbf{ARPES spectra of a (110) surface of LSCO.} \textbf{a}) Calculated three-dimensional Fermi surface of LSCO showing the contours at $k_{z} = 0$ (red) as in panel \textbf{b} and the nodal cut (blue) as in panel \textbf{c}. \textbf{b}) Varying the photon energy from a (110) surface yields a cut of the well known (001) Fermi surface of LSCO shown in the left plot. The right plot shows the calculated (001) Fermi surface at the $\Gamma X \Sigma$ plane, for comparison. \textbf{c}) On the other hand, varying the analyzer deflector angle allows to directly obtain the dispersion of the bands along the [001] direction. Here the photon energy was 79 eV, corresponding to the nodal cut shown also in the next panel. \textbf{d}) Dispersion of the bands at the superconducting nodes taken with a photon energy of 79 eV, and (\textbf{e}) at the antinodes taken with a photon energy of 95 eV. Notice that due to $k_{z}$ broadening, the small dispersion of the bands at the antinodes is not recognizable as the bands become smeared out along the [110] direction, along which they have a large dispersion. This results in the strong intensity observed between the two visible bands. All the calculations include a $k_{z}$ broadening of $\Delta k_{[110]} = 0.2\ \text{\AA}^{-1}$, based on the typical inelastic mean-free path of photoelectrons with energies in the Vacuum Ultraviolet (VUV) region, which is about $5\ \text{\AA}$. All measurements were performed at 6~K.}
    \label{figure_2}
\end{figure}

The quality of the cleaved surface enables us to acquire high-resolution ARPES spectra, which in turn allow for a complete mapping of the three-dimensional Brillouin zone of LSCO (Fig.~\ref{figure_2}). Due to the orientation of the cleaved side surface, varying the incident photon energy permits momentum scans along the [110] crystallographic direction (i.e. along $k_z$ perpendicular to the cleaved side surface), thereby recovering the well-known Fermi surface in the $(001)$ plane of LSCO (Fig.~\ref{figure_2}b). 
Conversely, a simple deflection scan of the electron analyzer provides direct access to the electronic structure along the plane spanned by the [001] and [1$\bar{1}$0] crystallographic directions (Fig.~\ref{figure_2}c). 
Finally, by selecting appropriate photon energies, we can resolve the electronic dispersions at both the nodal and antinodal regions of the Brillouin zone (Fig.~\ref{figure_2}d–e). It should be noted that, due to $k_{z}$ broadening, the weak dispersion of the bands at the antinodal region cannot be resolved. This effect is also evident in the tight-binding calculations (see the Methods section~\ref{Methods:TB_3D}), where the $k_{z}$ broadening was explicitly included. The inelastic mean free path of the photoemitted electrons was evaluated using the Tanuma–Powell–Penn (TPP-2M) formula~\cite{Tanuma_1991} (see Supplementary Section~\ref{Methods:TPP-2M}) and used to determine the magnitude of the $k_z$ broadening, yielding a value on the order of $0.4~\text{\AA}^{-1}$. We also noticed that the results of the calculations seem to slightly overestimate the amount of $k_{z}$ broadening. In particular in Fig.~\ref{figure_2}c the bands show some partial $k_{z}$ dependence around $k_{[001]} =0$ coming from the antinodal region. This indicates that the TPP-2M formula might be slightly underestimating the inelastic mean free path. The tight-binding model used here is from Ref.~\cite{Markiewicz_2005}, where the chemical potential was chosen so as to obtain the same value of $k_{F}$ measured experimentally in the nodal cut. On the other hand, the effective doping of the measured surface was estimated to be approximately $p = 0.22$ based on the area of the fitted Fermi pockets (see the Supplementary section~\ref{Supplementary:Luttinger}). Despite the limitations on the accuracy of this estimate, the obtained value (which for different fitting parameters and constraints was consistently in the range $x = 0.2-0.25$, and only occasionally reaching up to about $x = 0.3$) indicates little variation of the effective doping between the bulk and surface values, suggesting that the surface is unlikely to undergo a reconstruction as reported in other cuprate superconductors~\cite{Hossain_2008}.

\subsection{Broadening of the surface states and gap frustration by pair-breaking}

Although surface states are typically regarded as effectively two-dimensional states localized at the boundary of the system, this picture is modified in the limit of a small gap, as occurs in the case of topological superconductors. For comparison, typical chiral topological semimetals exhibit projected gaps on the order of 0.1-1 eV~\cite{Schroter_2019,Schroter_2020} leading to strong confinement of Fermi-arc surface states at the surface, whereas the magnitude of the superconducting gap of LSCO at the antinodes is typically $10\text{–}20~\mathrm{meV}$~\cite{Yoshida_2016,Zhong_2022}, consistent with our $(001)$-surface reference measurement on a comparably prepared sample (Supplementary section~\ref{Supplementary:SC_001}). In this regime, the surface states are no longer strongly confined to the boundary but instead acquire a finite degree of bulk character and the spectral weight develops a dependence on the $k_z$  momentum component perpendicular to the surface (Fig.~\ref{figure_3}a). This behavior can also be understood by considering the limit where the bulk superconducting gap approaches zero. In this case, the system should host only bulk states and the spectral weight associated with zero-energy modes should evolve continuously into the spectral weight of the normal-state Fermi surface bounded by the projection of the superconducting nodes (see the Supplementary section~\ref{Supplementary:OOP_momentum_Dependence}). From an experimental standpoint, one must then select regions of the Brillouin zone corresponding to $k_z$  in which the spectral weight of the bulk superconducting state deviates the most from the spectral weight of the surface state, while ensuring that the surface-state spectral function retains sufficient intensity.

For $d_{x^{2}-y^{2}}$-wave superconductors, this condition should be satisfied near the antinodes ($h\nu=95~\mathrm{eV}$, see Fig.~\ref{figure_2}e), where the opening of the superconducting gap would be expected to deplete the bulk states at the Fermi level. However, this gap actually gets filled in the near surface region, since  a $d_{x^{2}-y^{2}}$-wave order parameter is suppressed at a (110) surface due to the phase shift acquired by reflected quasiparticles, which in turn frustrates the order parameter over a length scale comparable to the superconducting coherence length (Fig. \ref{figure_3}b-c and Refs. \cite{Fogelstr_m_1997,Bruder_1990,Misra_2002}), which is on the order of a few nanometers in the cuprates~\cite{Hwang_2021,Mangel_2024,Petrenko_2022}, larger than the typical probing depth of ARPES (see the Methods section~\ref{Methods:TPP-2M}).  Therefore, one in general does not expect the superconducting gap to be observable at a (110) surface, which poses a potential complication for separating the edge and bulk spectral weight. Remarkably, despite the suppression of the superconducting gap, the simulated spectra corresponding to a $k_z$ value at the antinodes on the (110) surface still show a discernible dip and peak in the spectra (Fig. \ref{figure_3}d, black arrow in Fig. \ref{figure_3}e). This feature may serve as a characteristic signature for identifying the surface states irrespective of the gap suppression. Indeed, we find that suppressing the superconducting gap at the surface has only a minimal effect on the surface-state spectral function (see Supplementary Fig.~\ref{Supplementary:edge_w_and_wo_gap_suppression}).

Nevertheless, despite bulk superconductivity being confirmed in our samples by magnetic susceptibility measurements (Fig.\ref{figure_2_FIB}b), and the surface doping level consistent with the bulk value (see the Supplementary section~\ref{Supplementary:Luttinger}), our experimental data from $k_z$ values corresponding to the antinodes exhibit no clear dip and peak at the Fermi level (see Fig. \ref{figure_3}g and the corresponding integration ranges in the Supplementary section~\ref{Supplementary:EDC_boundary}). Furthermore, within our experimental energy resolution of $3\text{–}4~\mathrm{meV}$, we do not observe a superconducting gap, which is expected to be approximately $20~\mathrm{meV}$ at our doping level~\cite{Zhong_2022}. This absence may arise from pair-breaking induced by the (110) surface or from significant broadening of the surface states. Notice that the small temperature dependence of the intensity observed in both the band and Topological sector Energy Distribution Curves (EDCs) was not reproduced in another sample that was measured at different temperatures (Supplementary section~\ref{Supplementary:Additional_EDCs}), thus we attribute this to extrinsic effects such as minor thermal drift or variations in surface conditions. 

To account for these observations, we incorporated geometric edge disorder in our simulations, implemented by removing sites along the edge using the experimentally measured roughness profile (Fig.~\ref{figure_2_FIB}e) obtained by AFM. Owing to the single-orbital nature of our model, the effective spatial resolution along the [11] direction is limited to $\frac{a\sqrt{2}}{2}$, where $a$ is the lattice constant. The resulting simulated spectra still exhibit the dip and surface-state peak (Fig.~\ref{figure_3}e), since the disordered edges still exhibit sufficiently long clean segments extending over multiple coherence lengths (Fig.~\ref{figure_3}c, top panel). To achieve a substantial broadening of the surface states, the surface height must vary with very high in-plane spatial frequency---that is, the roughness must entirely eliminate any contiguous flat regions down to the size of a single unit cell. Such lateral resolution is beyond the capabilities of conventional AFM setups, which are typically limited to a few nanometers~\cite{Santos_2011, Binnig_1986, Zhou_2017}. Moreover, the small sample size (on the order of a few hundred micrometers) makes it difficult to reliably engage with a scanning tunneling microscopy (STM) tip. Nevertheless, our results establish an upper bound on the disorder autocorrelation length and show that, for all disorder resolvable by our AFM measurement, the surface states remain intact. Finally, disorder with sufficiently high in-plane spatial frequency would be expected to partially lift the suppression of the gap at the edge (Fig.~\ref{figure_3}b and orange arrow in Fig.~\ref{figure_3}e). Although the emergence of a fully developed gap would likely remain unobservable under these conditions, one may nevertheless expect subtle modifications of the low-energy spectral features in agreement with the closure of the gap upon varying the temperature across the critical temperature. However, we do not observe such signatures.



Beyond geometric disorder at the surface, it is well established that most high-temperature cuprate superconductors exhibit substantial intrinsic disorder, arising primarily from chemical doping and oxygen vacancies~\cite{Sulangi_2018,Chen_2012,Pasupathy_2008,Graser_2007,Tolpygo_1996}. While more realistic descriptions usually account for this disorder through a distribution of point-like and extended impurity potentials with characteristic strengths ranging from several tens to several hundreds of meV~\cite{_zdemir_2022,Lee_Hone_2020,Nunner_2006}, the exact microscopic details are not required for developing qualitative intuition regarding its impact on the surface states. Accordingly, consistent with the deliberately minimal character of the present model, we approximate the effects of bulk disorder by an Anderson-type formulation implemented as a random onsite mass term. Importantly, although disorder preserving chiral symmetry does not mix states with the same chirality eigenvalues, scattering between different chirality sectors is still allowed (Supplementary section \ref{Supplementary:Scattering}). Since eigenstates of opposite chirality reside at opposite momenta, this inter-sector coupling necessarily involves large-momentum-transfer scattering processes, rendering the surface states particularly susceptible to short-range disorder, which is therefore expected to endow the flat band with a finite energy bandwidth proportional to the disorder strength. Indeed, we find that the inclusion of a random mass term at each site, with a magnitude drawn from a Gaussian distribution with a standard deviation of $100~\mathrm{meV}$ is able to entirely lift the degeneracy of the flat band in our simulations, thus strongly suppressing the peak from the flat band (Fig.~\ref{figure_3}d-e). We further observe a relative increase in spectral weight in the trivial sector near the Fermi level (Fig.~\ref{figure_3}e); however, this feature may be experimentally challenging to resolve due to background intensity and the possible contribution of geometric disorder with higher in-plane spatial frequencies. The simulated cut at the antinodes with both edge roughness and onsite disorder is shown in the right half of Fig.~\ref{figure_3}d, highlighting the visual agreement with the experimental results (Fig.~\ref{figure_3}f). Notice that this amount of disorder is not inconsistent with the observation of a superconducting  gap at surfaces where there is no surface state, such as the (001) surface which is typically measured by ARPES experiments (Supplementary section \ref{Supplementary:gap_001_GE+BA}).

\subsection{Discussion}

\begin{figure}[H]
    \centering
    \includegraphics[width = 1\textwidth]{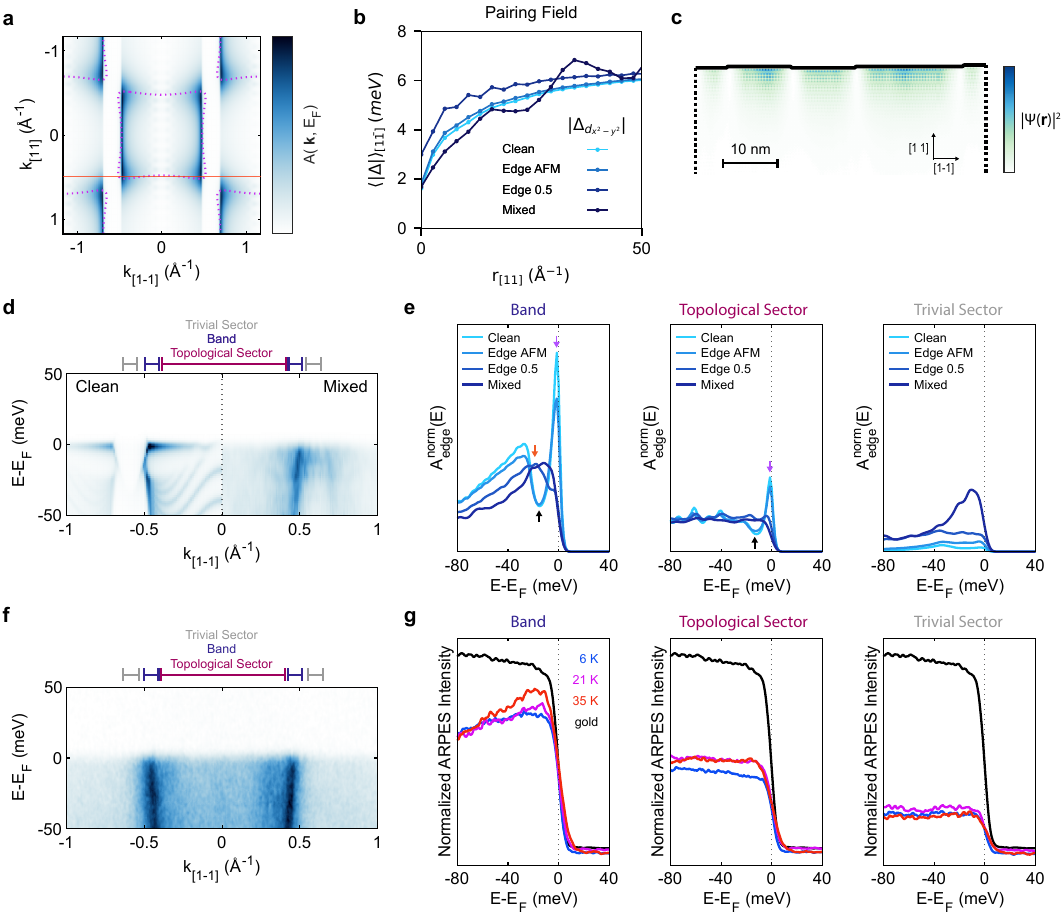}
   \caption{\textbf{Boundary effects at a (110) surface of a $d_{x^{2}-y^{2}}$-wave superconductor.}\textbf{a}) Simulated spectral function of the [11] edge at the Fermi level. \textbf{b}) $d_{x^{2}-y^{2}}$ pairing amplitude versus distance from the [11] edge for a clean system (C), AFM-derived geometric edge roughness (GE${\mathrm{AFM}}$), probabilistic roughness (GE${0.5}$), and GE${\mathrm{AFM}}$ combined with bulk Anderson disorder (BA). The BA disorder is Gaussian with a standard deviation of \(100~\mathrm{meV}\). \textbf{c}) Real-space expectation value of edge-state eigenstates for AFM-derived geometric edge roughness. States are selected by their eigenvalues and participation ratios. Dotted lines indicate periodic boundary conditions. \textbf{d}) Simulated antinodal spectral functions for C (left) and GE$_{\mathrm{AFM}}$ + BA disorder (right), extracted along the red line in \textbf{a}. \textbf{e}) Simulated antinodal EDCs for different disorder realizations and $k{[11]}$; purple, black, and orange arrows mark the surface-state peak, DOS dip, and coherence peak, respectively. \textbf{f}) ARPES spectra measured with 95~eV photons and at 6~K (see Fig.~\ref{figure_2}b). \textbf{g}) Antinodal EDCs (\(h\nu = 95~\mathrm{eV}\)) normalized by total intensity at different temperatures, compared with the gold DOS obtained at 6~K. The integration ranges are indicated by the colored lines above \textbf{d} and \textbf{f} (see also the Supplementary section~\ref{Supplementary:EDC_boundary}). Reported values are mean intensities within these ranges, normalized to the total intensity. For the ARPES data, a background subtraction following Ref.~\cite{Matt_2018} was applied prior to the EDC extraction. All calculations are performed at \(0~\mathrm{K}\).
}
    \label{figure_3}
\end{figure}

With this interpretation, the ZBCPs observed in STM measurements~\cite{Sharoni_2001,Misra_2002,Alff_1997} can plausibly be attributed to the differences in spatial resolution compared to ARPES, which is limited by the beam spot size (approximately $50~\mathrm{\mu m}$ in our measurements), so that one cannot isolate nanoscale regions with lower disorder. Indeed, as shown in Fig.~\ref{figure_3}c, the edge roughness already produces an inhomogeneous spatial distribution of the edge state wavefunction. Upon introducing Anderson disorder, this spatial fragmentation is further enhanced, such that the surface states may reside only on isolated segments of the boundary whose contribution to the photoemission intensity is strongly diluted when averaging over the full beam spot area. This effect has also been reported in STM measurements at step edges~\cite{Misra_2002}, where the relative prominence of the zero-bias peak was found to vary substantially between different positions along the edge. Crucially, the inclusion of on-site disorder enhances this spatial inhomogeneity, further complicating the isolation of the edge state by ARPES. Even when a relatively uniform ZBCP was observed in STM measurements~\cite{Alff_1997}, the reported spatial range only covers about $16~\mathrm{nm}$, which is two orders of magnitude smaller than our spatial resolution. 

On the other hand, tunnel junction measurements, which also probe a larger area but do not require an atomically smooth surface, still observe a zero-bias peak. One possible explanation is that dilute disorder can displace the surface states away from the topmost layer~\cite{Queiroz_2016}, reducing their visibility in surface-sensitive probes like VUV ARPES while allowing them to contribute to the tunneling signal. An additional explanation is that at least some of the ZBCPs observed in tunneling measurements could be due to interface states arising between domains with different orientations~\cite{Hu_1998}.
In the light of this complicated picture, one must notice that disorder can manifest in various forms which can have markedly different effects on the surface states depending on their symmetries as well as their spatial distribution. For instance, one could consider onsite disorder, such as chemical potential disorder or mass disorder (used in our calculations), as well as disorder in the hopping terms. These types of disorder can furthermore be either localized near the edge or extended throughout the bulk, and their distribution can, for instance, be Gaussian or dilute~\cite{Queiroz_2016}, and exhibit varying degrees of spatial correlation. Determining which type of disorder affects a certain measurement is thus evidently challenging in an experimental setting.

Our work however, highlights that this new cleaving technique is capable of producing surfaces with a topographic roughness that, within our experimental resolution and according to our microscopic calculations, is not sufficient to significantly lift the degeneracy of the surface states, thus highlighting the applicability of the technique to quasi-two-dimensional systems. Furthermore, from a measurement perspective, it indicates that the main challenge in detecting momentum resolved signatures of these flat band surface states by ARPES lies in the ability to resolve states characterized by a strong spatial inhomogeneity, which under certain circumstances might be pushed away from the topmost surface layers. On the other hand, from a material point of view, our results demonstrate the importance of minimizing bulk disorder, for instance by studying materials with low defect densities and near-stoichiometric compositions. Thus, our results lay essential groundwork for future investigations of topological superconductors using ARPES, while also demonstrating how to study materials with surface sensitive probes for which conventional mechanical cleavage is challenging.

\newpage

\section{Methods}

\subsection{Sample Preparation}

Single crystals were grown by the traveling solvent floating-zone method. Crystal growth was performed under an oxygen pressure of $2~\mathrm{amt}$. Post-growth annealing was conducted at $800~\mathrm{^{\circ}C}$ in an oxygen atmosphere for $100~\mathrm{h}$. The Strontium concentration was determined from the critical temperature~\cite{Momono_2002,Matsuzaki_2004}.
Small pillars with approximate dimensions of $0.5 \times 0.5 \times 1.5~\mathrm{mm^3}$ were cut from the resulting rod, with the longest axis aligned along the desired cleavage direction. 
To minimize the contact resistance between the sample and the sample stub, a thin layer of gold was deposited on the surface of the pillars, followed by an annealing treatment in an oxygen atmosphere at $1~\mathrm{bar}$ for $1~\mathrm{h}$ at $600^{\circ}\mathrm{C}$~\cite{Ekin_1988}. The samples were then mounted on copper stubs and fixed in place using EPO-TEK\textsuperscript{\textregistered} H21D silver epoxy, which was cured at $80^{\circ}\mathrm{C}$ for $90~\mathrm{min}$. 
The stub was subsequently mounted on a custom-made SEM extender, allowing for a manipulator tilt of up to $-35^{\circ}$. This geometry enabled perpendicular milling of all sides of the pillar without the need for repeated sample removal and remounting. To minimize possible local heating during the milling process, two strategies were employed: (\textit{i}) milling was performed in parallel on opposite sides of the pillar (i.e., alternating milling cycles between opposing faces), and (\textit{ii}) the maximum beam current was restricted to $500~\mathrm{nA}$ and progressively reduced down to $1~\mathrm{nA}$ as the cross-section of the pillar was thinned to approximately $150 \times 150~\mathrm{\mu m^2}$.
The comparison of the superconducting critical temperature, obtained from the temperature dependence of the magnetic susceptibility of both the as-grown floating-zone rod and a small milled portion of the sample, revealed no significant change in onset temperature or overall behavior (Fig.~\ref{figure_2_FIB}b). 
Finally, the copper stub was directly mounted onto an ARPES sample plate, prepared for in-situ mechanical cleaving achieved by gently applying side pressure to the top of the pillar. 
\subsection{ARPES}

The ARPES measurements were performed at the I05 beamline of Diamond Light Source~\cite{Hoesch_2017} with a MBS A-1 hemispherical analyzer (MB Scientific AB) under an ultrahigh-vacuum pressure of approximately $1\text{–}2 \times 10^{-10}~\mathrm{mbar}$. The photon energies used were $h\nu=79~\mathrm{eV}$ for the nodal cut, $h\nu=95~eV$ for the antinodal cut and in the range of $h\nu=40–200\ eV$ for the photon energy dependence maps. The sample temperature ranged from $6~\mathrm{K}$ to $35~\mathrm{K}$. The combined energy resolution was approximately $3\text{–}4~\mathrm{meV}$ with an approximate beam spot diameter (FWHM) of $40\text{–}50~\mathrm{\mu m}$.

\subsection{\label{Methods:TB_3D}Three-dimensional Fermi surface}

The simulated three-dimensional Fermi surface of LSCO was modelled by a one-band tight-binding model~\cite{Markiewicz_2005} where we have set the chemical potential $\mu = -393.3~\mathrm{meV}$ in order to match the value of $k_{F} = -0.4605\ \text{\AA}$ obtained experimentally in the nodal cut.

\subsection{\label{Methods:TPP-2M}Tanuma-Powell-Penn formula}

The TPP-2M formula~\cite{Tanuma_1991} allows to obtain an estimate of the inelastic mean free path of photoemitted electrons based on a series of materials parameters. Using a density of $7\ g/cm^{3}$ and a band gap of $20~\mathrm{meV}$, corresponding to the superconducting gap at the antinodes for $x = 0.22$~\cite{Zhong_2022}, we obtain an inelastic mean free path $\lambda = 4.7413\ \text{\AA}$, that is in agreement with published estimates~\cite{Damascelli_2003}. The Fourier transform of an exponentially decaying function with a characteristic length scale $\lambda$ is a Lorentzian with full width at half maximum equal to $2/\lambda$. This allows us to obtain an estimate for the momentum broadening of $\Delta k_{z} \approx 0.4\ \text{\AA}^{-1}$.

\subsection{\label{Methods:TB_2D}Two-dimensional slab}

In order to simulate the spectral function of the edge state, we constructed a single-orbital tight-binding model on a square lattice geometry defined over a cylindrical slab. Specifically, we imposed periodic boundary conditions along one spatial direction and open boundary conditions along the orthogonal direction, thereby generating two opposing edges oriented along the $[11]$ crystallographic direction. The cylindrical slab comprised $280$ sites in the direction perpendicular to the edge and $200$ sites in the direction parallel to the edge.
The Hamiltonian incorporated hopping terms up to the third nearest neighbor, with the corresponding hopping amplitudes obtained from previously published fitting results to ARPES data on samples with the same doping levels~\cite{Yoshida_2007}. Furthermore, we used a nearest-neighbor pairing field whose local value was determined self-consistently (see the Methods section~\ref{Methods:Self_Consistency}). Random on-site disorder was introduced through a mass term, given in the Nambu basis by
\begin{equation*}
    V_i \;=\; v_i \, \sigma_z,
\end{equation*}
where $\sigma_z$ denotes the corresponding Pauli matrix and $v_i$ is a site-dependent random variable drawn from a Gaussian distribution with a standard deviation equal to the disorder strength.  
The full Hamiltonian was then diagonalized in real space and to account for the surface sensitivity inherent to ARPES measurements, the resulting eigenstates were weighted by an exponential function decaying into the bulk with a characteristic length scale equal to the inelastic mean free path of photoemitted electrons, which was estimated using the TPP-2M formula (see the Methods section~\ref{Methods:TPP-2M}). Notice that because the lattice spacing for the discrete Fourier transform is comparable to the inelastic mean free path, some degree of aliasing artifacts is expected. Moreover, the finite size of the slab acts as a rectangular window introducing spectral leakage. Both of these effects can result in a partial overestimation of the effective $k_{z}$ broadening. Finally, in order to access the full momentum dependence probed by ARPES, we applied a two-dimensional Fourier transform to the exponentially weighted eigenstates -- hence directly including a $k_{z}$ broadening -- and computed the momentum dependent spectral function~\cite{Schubert_2012, Queiroz_2015}, where in the retarded Green's function we used a broadening of $3~\mathrm{meV}$ to match the experimental energy resolution.

\subsection{\label{Methods:Self_Consistency}Self consistent Bogoliubov-de Gennes gap equation}

The d-wave superconductivity can be captured within a mean-field treatment of the t-J model, where the quartic interaction term --- omitting the density term --- is decoupled in the pairing channel~\cite{Sachdev_2023}

\begin{equation}
\mathcal{H}_J = J\sum_{\braket{ij}} \mathbf{S}_{i} \cdot \mathbf{S}_{j} = -\frac{J}{2}\sum_{\braket{ij}}\Delta_{ij}\varepsilon_{\alpha\beta}c^{\dagger}_{i,\alpha}c^{\dagger}_{j,\beta} + \Delta^{*}_{ij}\varepsilon_{\alpha\beta}c_{j,\beta}c_{i,\alpha} - |\Delta_{ij}|^{2},
\label{eq:quartic_term_spinful}
\end{equation}

where $\varepsilon_{\alpha\beta}$ is the Levi-Civita symbol with $\varepsilon_{\uparrow\downarrow} = 1$, and the spin-singlet pairing amplitude is defined as

\begin{equation}
\Delta_{ij} = - \braket{c_{i,\uparrow}c_{j,\downarrow} - c_{i,\downarrow}c_{j,\uparrow}}.
\label{eq:self_consistent_relation_spinful}
\end{equation}

In the absence of spin-orbit coupling, the system preserves SU(2) spin symmetry, and the Hamiltonian decouples into two identical singlet pairing blocks

\begin{equation}
\mathcal{H}_{J} = \mathcal{H}_{\uparrow \downarrow} \oplus \mathcal{H}_{\downarrow \uparrow}.
\label{eq:direct_sum}
\end{equation}

Consequently the problem can be reformulated in a reduced Nambu basis, where for the Nambu spinor \[
\Psi_i = 
\begin{pmatrix}
c_{i \uparrow} \\
c_{i \downarrow}^\dagger
\end{pmatrix},
\] Eqs.~\ref{eq:quartic_term_spinful} and Eqs.~\ref{eq:self_consistent_relation_spinful} are:

\begin{equation}
\mathcal{H}_J = J\sum_{\braket{ij}} \mathbf{S}_{i} \cdot \mathbf{S}_{j} = -\frac{J}{2}\sum_{\braket{ij}}\Delta_{ij}c^{\dagger}_{i\uparrow}c^{\dagger}_{j \downarrow} + \Delta^{*}_{ij}c_{j \downarrow}c_{i  \uparrow} - |\Delta_{ij}|^{2},
\label{eq:quartic_term_spinless}
\end{equation}

and 

\begin{equation}
\Delta_{ij} = - \braket{c_{i \uparrow}c_{j \downarrow}}.
\label{eq:self_consistent_relation_spinless}
\end{equation}

Neglecting the effects of the projection on the Hilbert space without doubly occupied sites, as well as the corrections to the kinetic terms and the chemical potential, we only need to impose self-consistency of the pairing gap~\cite{Chakraborty_2022}, which is done by solving the Bogoliubov-de Gennes self-consistent gap equation~\cite{Bogoljubov_1958,DE_GENNES_1964}

\begin{equation}
\Delta_{ij} \;=\; V \sum_{n} 
u_n(i)\, v_n^*(j)\, \tanh\!\left(\frac{E_n}{2 k_B T}\right),
\label{eq:gap_self_consistent}
\end{equation}

Where $V$ denotes the effective pairing interaction, $u_{n}(i)$ and $v_{n}(i)$ are the local amplitudes of the Bogoliubov quasiparticles and $E_{n}$ are their corresponding eigenvalues.
The value of $V = 0.153~\mathrm{meV}$ ($V = 0.21555~\mathrm{meV}$ when the onsite disorder is introduced) was chosen so as to get a gap of approximately $24~\mathrm{meV}$ at the antinodes, in approximate agreement with experimental values~\cite{Zhong_2022}.

To improve the convergence to the fixed point solution we implemented an Anderson-Pulay method~\cite{Chupin_2021} with $m = 5$, combined with a mixing step 

\begin{equation}
\Delta_{k+1}^{(mix)} = \left(1 - \beta\right) \Delta_{k} + \beta \Delta_{k+1}^{(new)},
\label{eq:mixing_step}
\end{equation}
where $k$ is the iteration index and with $\beta = 0.7$ for the calculations with onsite disorder and $\beta = 0.8$ otherwise.

\section{Data Availability}

\section{Acknowledgments}

G. D., M. D., S.K.Y.D, N. L., A. K. and D. P acknowledge proposals SI35157, SI35176 and SI36075 at the i05 Endstation at the Diamond Light Source, UK. G. D. acknowledges the Institute for Factory Automation and Production Systems for the access to their PFIB funded by the Deutsche Forschungsgemeinschaft (DFG, German Research Foundation) – 442921285. A.P.S. is funded by the Deutsche Forschungsgemeinschaft (DFG, German Research Foundation) – TRR 360 – 492547816. G.D. acknowledges support by the Max Planck Graduate Center for Quantum Materials (MPGC-QM). T. K. acknowledges funding by the JSPS KAKENHI Grant No. 22K03504. G.D. and S. T. acknowledge proposal 20227155 at the URANOS beamline at the National Synchrotron Radiation Centre “SOLARIS”, Poland. This publication was partially developed under the provision of the Polish Ministry of Science and Higher Education project 'Support for research and development with the use of research infrastructure of the National Synchrotron Radiation Centre SOLARIS' under contract no 1/SOL/2021/2. N.B.M.S. acknowledges funding by the European Union (ERC Starting Grant ChiralTopMat, Project No. 101117424). We acknowledge Laura Green, Hideaki Iwasawa, and Bernhard Keimer for helpful discussions about this topic. G. D. wishes to express his appreciation to Debmalya Chakraborty for the many insightful discussions.

\section{Supplementary}

\subsection{\label{sup_sub_sec_1}Analytical expression of the edge-state wave function}

An effectively single-band d-wave superconductor belongs to the symmetry class CI. Since it is a gapless system, it is not possible to define a global topological invariant, however, away from the nodes, the presence of the chiral symmetry allows to define a group of maps $d_{k}: S^1 \mapsto \mathbb{R}\setminus\{0\} \simeq S^1$. For non-contractable loops $k \in S^{1} \subseteq T^{2}$, the equivalence classes of the maps form the fundamental homotopy group $\pi_{1}\left(\mathbb{R}\setminus\{0\}\right) = \mathbb{Z}$, where each class can be labeled by a winding number. A non-zero value of this topological invariant along a path with constant momentum parallel to an edge, denoted as $k_{\parallel}$, implies the presence of a zero-energy edge state with the same momentum $k_{\parallel}$~\cite{Ryu_2002,Sato_2011}. 

\begin{figure}[h!]
    \centering
    \includegraphics[width = 0.7\textwidth]{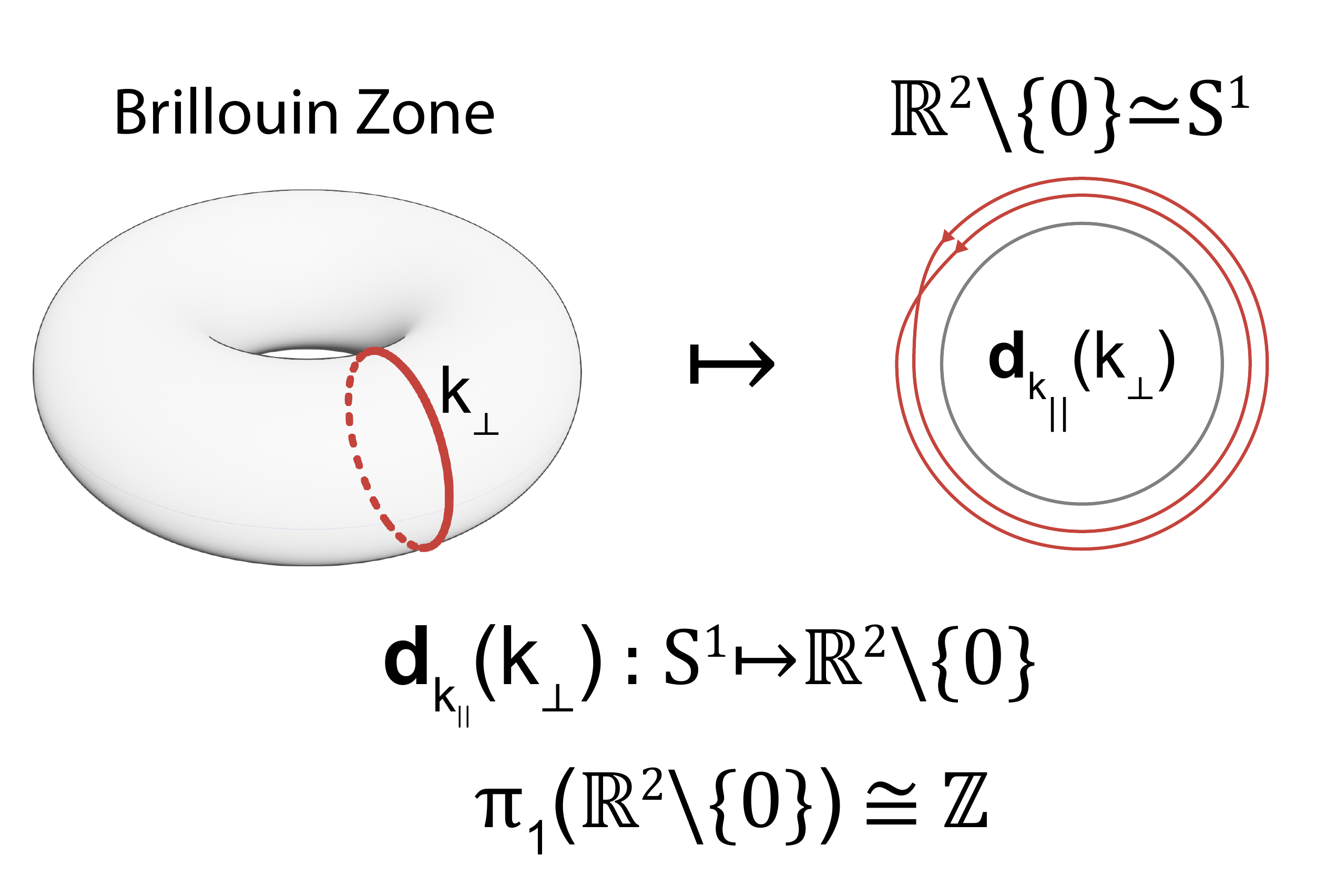}
    \caption{Let $X \coloneq [-\pi, \pi] \times [-\pi, \pi]$ be a square Brillouin zone with the equivalence relation $(k_{\parallel},-\pi) \sim (k_{\parallel},\pi)$ and $(-\pi,k_{\perp}) \sim (\pi,k_{\perp})$. Then the quotient space $X/\sim$ is a $T^{2}$ torus. Then paths with a constant momentum component are topologically equivalent to non-contractable $S^{1}$. Each map $d_{k}$ is then connecting the elements of these loops to the elements of another $S^{1}$ loop.}
    \label{fig:BZ}
\end{figure}

A single-band class CI superconductor can be described by the following Bogoliubov-de Gennes (BdG) Hamiltonian:

\begin{align}
\mathcal{H}(\bm{k}) =  \frac{1}{2}\sum_{\bm k}(c_{\bm k \uparrow}^{\dagger},c_{\bm{-k} \downarrow})\mathcal{H}_{BdG}(\bm{k})
\left(
\begin{array}{c}
c_{\bm k \uparrow} \\
c_{\bm {-k} \downarrow}^{\dagger}
\end{array} \right)
\label{eq:total_hamiltonian}
\end{align}

\begin{align}
\mathcal{H}_{BdG}(\bm k) = 
\left(
\begin{matrix}
\varepsilon_{\bm{k}} & \Delta_{\bm{k}}\\
\Delta_{\bm{k}} & -\varepsilon_{\bm{k}}
\end{matrix}\right)\;
\label{eq:BdG_hamiltonian}
\end{align}

Where for a square lattice belonging to the two-dimensional point group $D_{4}$ we have:

\begin{align}
\varepsilon_{\bm{k}} = -4t \cos\left(k_{\parallel} \frac{\delta}{2}\right) \cos\left(k_{\perp} \frac{\delta}{2}\right) - \mu
\label{eq:BdG_hamiltonian_epsilon}
\end{align}

\begin{align}
\Delta_{\bm{k}} = -4|\Delta|\sin\left(k_{\parallel} \frac{\delta}{2}\right) \sin\left(k_{\perp} \frac{\delta}{2}\right)
\label{eq:BdG_hamiltonian_delta}
\end{align}

Here, for convenience, we have rotated the square lattice so that $\delta = a\sqrt{2}$, with $a$ being the lattice spacing between two nearest neighbours.

By construction, this BdG Hamiltonian has a spinless time-reversal symmetry $\mathcal{T} = \mathcal{K}$, where $\mathcal{K}$ is the complex conjugation operator, and a charge conjugation or particle-hole symmetry $\mathcal{C} = \sigma_{y}\mathcal{K}$. Combining these two symmetries we additionally obtain a chiral symmetry which anticommutes with the BdG Hamiltonian:

\begin{align}
    \{ \Gamma , \mathcal{H}_{BdG} (\bm{k}) \} = 0 \;, \;\;\;\;\Gamma = \sigma_{y}
\label{eq:Chiral_operator}
\end{align}

By diagonalizing the chiral operator we can obtain the unitary matrix $U_{\Gamma}$ that maps an operator to its chiral basis representation.

\begin{align}
    U_{\Gamma}^{\dagger} \Gamma U_{\Gamma} = \sigma_{z} \;, \;\;\;\; U_{\Gamma} = \frac{1}{\sqrt{2}} \left(
    \begin{matrix}
    1 & 1\\
    i & -i
    \end{matrix}\right)
\label{eq:Chiral_operator}
\end{align}

So that we can bring $\mathcal{H}_{BdG}$ into the following off-diagonal form:

\begin{align} 
    \mathcal{H}_{BdG}^{(\Gamma)}(\bm{k}) &= U_{\Gamma}^{\dagger} \mathcal{H}_{BdG}(\bm{k}) U_{\Gamma}
    \label{eq:pseudospin} \\ &= \left(
    \begin{matrix}
    0 & q_{\bm{k}}\\
    q^{*}_{\bm{k}} & 0
    \end{matrix}\right) \notag \\ 
    &= \left(
    \begin{array}{c}
    \mathfrak{R}(q_{\bm{k}}) \\
    -\mathfrak{I}(q_{\bm{k}})
    \end{array}\right) \cdot \bm{\sigma} \notag \\
    &= \bm{d}_{\bm{k}} \cdot \bm{\sigma}  \notag 
\end{align}

Where $q_{\bm{k}} = \varepsilon_{\bm{k}} - i \Delta_{\bm{k}}$

This simply amounts to perform a change of basis from the particle-hole basis to the chiral basis:

\begin{align}
    U^{\dagger}_{\Gamma} \left(
    \begin{array}{c}
    c_{\bm{k} \uparrow} \\
    c_{\bm{-k} \downarrow}^{\dagger}
    \end{array} \right) &= \frac{1}{\sqrt{2}}\left(
    \begin{array}{c}
    c_{\bm k \uparrow}-ic_{\bm{-k} \downarrow}^{\dagger} \\
    c_{\bm k \uparrow}+ic_{\bm{-k} \downarrow}^{\dagger}
    \end{array} \right) \notag \\ 
    &= \left(
    \begin{array}{c}
    \gamma_{\bm{k},+} \\
    \gamma_{\bm{k},-}
    \end{array} \right) = \bm{\gamma}_{\bm{k}}
\label{eq:chiral_bogoliubov}
\end{align}
Where $\gamma_{\bm{k},+}$ and $\gamma_{\bm{k},-}$ are annihilation operators of quasiparticles with chirality +1 and -1, respectively.

Using \ref{eq:pseudospin} and \ref{eq:chiral_bogoliubov}, the Hamiltonian of \ref{eq:total_hamiltonian} can be rewritten as:

\begin{align}
\mathcal{H}(\bm{k}) =  \frac{1}{2}\sum_{\bm k}\bm{\gamma}_{\bm{k}}^{\dagger}\mathcal{H}_{BdG}^{(\Gamma)}(\bm{k})
\bm{\gamma}_{\bm{k}}\; .
\label{eq:total_hamiltonian_chiral_basis}
\end{align}

We now want to use this result to get an expression for the in-gap states wave function. To do so, we begin by noting that the eigenvalue equation $\mathcal{H}_{BdG}\Psi_{k_{\parallel}} = E_{BdG}\Psi_{k_{\parallel}}$, for energies satisfying $|E_{BdG}| < \Delta_{\bm{k}}$,  only allows for exponentially decaying solutions. These eigenstates are obtained by allowing the crystal momentum to take on complex values.
Thus, we can devise the following ansatz consisting of a linear superposition of plane waves that decay exponentially into the bulk with a characteristic lengthscale $\xi = \frac{1}{\mathfrak{R}\left(\kappa^{\gamma\eta}\right)}$:

\begin{align}
\Psi_{k_{\parallel},-i\kappa_{m}^{n}}(r_{\parallel},r_{\perp}) =  \sum_{n,m} C^{n}_{m} \psi^{n}_{m} e^{\kappa_{m}^{n} r_{\perp}} e^{i k_{\parallel} r_{\parallel}}
\; .
\label{eq:ansatz}
\end{align}

Because the flat-bands are pinned at the Fermi level, we can get the allowed values for $\kappa$ by solving the secular equation $\mathcal{H}_{BdG}^{(\Gamma)}(k_{\parallel},-i\kappa_{m}^{n})\Psi_{k_{\parallel}}(r_{\parallel},r_{\perp}) = 0$, where we have used the momentum operator $-i\partial_{r}$. 
After expanding the trigonometric functions to second order in $\bm{k}$, we get two solutions for $q_{\bm{k}}$  and two solutions for $q^{*}_{\bm{k}}$:

\begin{align}
\kappa^{\gamma\eta} = -\gamma\frac{|\Delta|}{t} k_{\parallel} + (-1)^{\eta+1} \frac{1}{t} \sqrt{\left( |\Delta|^2 +t^2\right){k_{\parallel} }^2 -\frac{t(4t+\mu)}{a^2}}
\label{eq:kappa}
\end{align}

\begin{figure}[h!]
\includegraphics[width=0.2\textwidth]{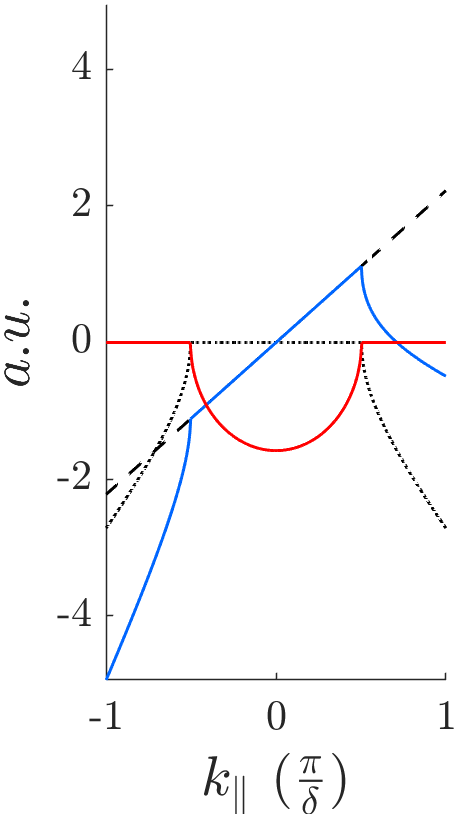}
\includegraphics[width=0.2\textwidth]{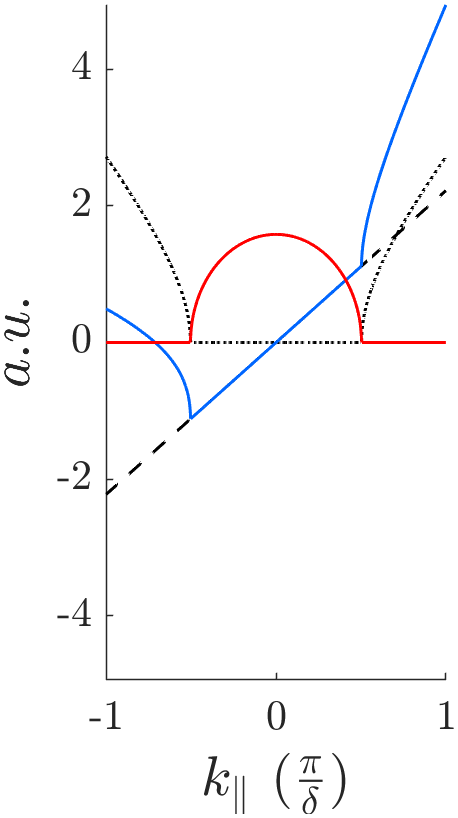}
\includegraphics[width=0.2\textwidth]{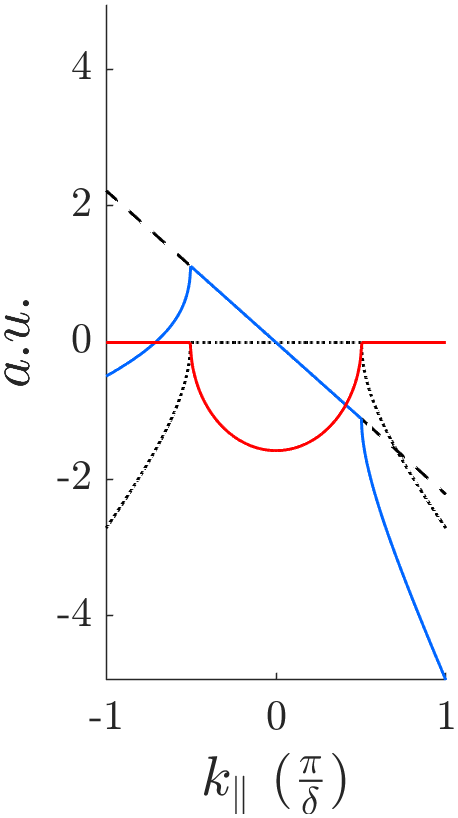}
\includegraphics[width=0.2\textwidth]{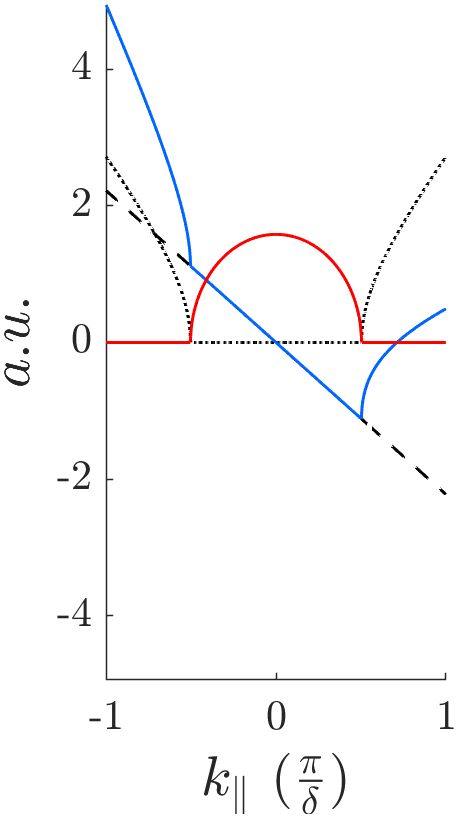}
\caption{\label{fig:kappa}Real (blue) and imaginary parts (red) of the four allowed $\kappa^{\gamma\eta}$ for $|\Delta|=t$. The dashed and dotted lines show the two contributions to the real part, coming respectively from the first and second term in \ref{eq:kappa}.}
\end{figure}

Where $\gamma \in \{ +,-\}$ and $\eta \in \{ 1,2\}$. The four solutions are plotted in Figure \ref{fig:kappa}. Here $\gamma$ indicates the eigenvalues of the chiral symmetry operator, since $q_{\bm{k}}$ and $q^{*}_{\bm{k}}$ respectively yield the solutions with chirality $-1$ and $+1$, spanned by $\gamma_{\bm{r},-}^{\dagger}\ket{0}$ and $\gamma_{\bm{r},+}^{\dagger}\ket{0}$ according to $\Gamma \gamma_{\bm{r},\pm}^{\dagger}\ket{0} = \pm\gamma_{\bm{r},\pm}^{\dagger}\ket{0}$. This leads to:


\begin{align}
\Psi_{k_{\parallel}}(\bm{r}) &=  \sum_{\gamma,\eta} C_{k_{\parallel}}^{\gamma\eta}  e^{\bm{K} \cdot \bm{r}} \gamma_{\bm{r},\gamma}^{\dagger}\ket{0} \\&= \sum_{\gamma} \varphi^{\gamma}_{k_{\parallel}}\gamma_{\bm{r},\gamma}^{\dagger}\ket{0}.
\label{eq:ansatz_2}
\end{align}

Where:

\begin{gather}
\bm{K} = \begin{pmatrix}K_{\parallel}\\K_{\perp}\end{pmatrix} = \begin{pmatrix}0\\\mathfrak{R}\left(\kappa^{\gamma\eta}\right)\end{pmatrix} + i \begin{pmatrix} k_{\parallel} \\ \mathfrak{I}\left(\kappa^{\gamma\eta}\right)\end{pmatrix} \; ,\\ \bm{r} = \begin{pmatrix}r_{\parallel}\\r_{\perp}\end{pmatrix}.
\end{gather}

To determine the coefficients $C^{\gamma \eta}_{k_{\parallel}}$ we focus on a semi-infinite system with $r_{\perp}>0$ and an edge at $r_{\perp}=0$, upon imposing that  $\Psi_{k_{\parallel},-i\kappa^{\gamma\eta}}(r_{\parallel},r_{\perp}=\infty) = 0$ we find the following conditions:

\begin{equation}
    \begin{cases}
        C_{k_{\parallel}}^{-\eta} = 0 & k_{\parallel} \in [-k_{F},0]\\
        C_{k_{\parallel}}^{+\eta} = 0 & k_{\parallel} \in [0,k_{F}]
    \end{cases}
    \label{eq:cases1}
\end{equation}

Furthermore the system has a time reversal symmetry, which in the chiral basis takes the form $\mathcal{T}_{\Gamma} = U_{\Gamma}^{\dagger} \mathcal{T} U_{\Gamma} = \sigma_{x} \mathcal{K}$. 
After applying this operator to the wave function, we get the conditions:

\begin{equation}
    \begin{cases}
        \varphi^{+}_{k_{\parallel}} = \left(\varphi^{-}_{k_{\parallel}}\right)^{*}\\
        \varphi^{-}_{k_{\parallel}} = \left(\varphi^{+}_{k_{\parallel}}\right)^{*}
    \end{cases}
    \label{eq:cases2}
\end{equation}

Lastly, the semi-infinite system also has a mirror plane $\sigma_{v} = \sigma_{x}\mathcal{K}_{\parallel}$ perpendicular to the edge, where the subscript denotes the fact that the complex conjugation operator acts only on $K_{\parallel}$. So that we get:

\begin{equation}
    \begin{cases}
        \varphi^{+}_{k_{\parallel}} = \mathcal{K}_{\parallel}\varphi^{-}_{k_{\parallel}}\\
        \varphi^{-}_{k_{\parallel}} = \mathcal{K}_{\parallel}\varphi^{+}_{k_{\parallel}}
    \end{cases}
    \label{eq:cases3}
\end{equation}

Combining \ref{eq:cases1}, \ref{eq:cases2} and \ref{eq:cases3} we obtain:

\begin{align}
   \sin\left(\mathfrak{I}\left(\kappa^{\gamma 1} \right) \right)\left(C^{\gamma 1}_{k_{\parallel}}e^{\mathfrak{R}\left(\kappa^{\gamma 1} \right)}-C^{\gamma 2}_{k_{\parallel}}e^{\mathfrak{R}\left(\kappa^{\gamma 2} \right)}\right) = 0.
\end{align}

Where we used the fact that $\mathfrak{I}\left(\kappa_{1}^{+} \right) = (-1)^{\alpha+1}\mathfrak{I}\left(\kappa^{\gamma \eta} \right)$. Because $\mathfrak{I}\left(\kappa^{\gamma 1}\right) = 0$ in the interval $k_{\parallel}^{*} = \sqrt{\frac{2t\left(4t+\mu\right)}{\delta^2 \left(|\Delta|^2+t^{2} \right)}} \leq |k_{\parallel}| \leq  \sqrt{\frac{2 (4t+\mu)}{t\delta^2}} = k_{F}$, the equation is always verified in this range. On the other hand, when $\mathfrak{I}\left(\kappa^{\gamma 1}\right) \neq 0$ we have that $\mathfrak{R}\left(\kappa^{\gamma 1}\right) = \mathfrak{R}\left(\kappa^{\gamma 2}\right)$, so that $C^{\gamma 1}_{k_{\parallel}} = C^{\gamma 2}_{k_{\parallel}}$. We can then impose that the wave function be continuous and differentiable at $k_{\parallel}^{*}$:

\begin{gather}
    \varphi^{\gamma}_{|k_{\parallel}|<k_{\parallel}^{*+}}\Bigr|_{\substack{k_{\parallel}=k_{\parallel}^{*-}}} =  \varphi^{\gamma}_{|k_{\parallel}|>k_{\parallel}^{*+}}\Bigr|_{\substack{k_{\parallel}=k_{\parallel}^{*+}}}\\
   \frac{d}{dk_{\parallel}} \varphi^{\gamma}_{|k_{\parallel}|<k_{\parallel}^{*+}}\Bigr|_{\substack{k_{\parallel}=k_{\parallel}^{*-}}} = \frac{d}{dk_{\parallel}} \varphi^{\gamma}_{|k_{\parallel}|>k_{\parallel}^{*+}}\Bigr|_{\substack{k_{\parallel}=k_{\parallel}^{*+}}},
\end{gather}

which ultimately requires that $C^{\gamma 1}_{k_{\parallel}} = C^{\gamma 2}_{k_{\parallel}}$ be verified for all $k_{\parallel}$. Finally by using again \ref{eq:cases3} we obtain that:

\begin{align}
   C_{k_{\parallel}} = C^{+ \eta}_{k_{\parallel}} =  C^{- \eta}_{-k_{\parallel}}.  \rlap{$\qquad \Box$}
\end{align}

Because for each state identified by $k_{\parallel}$ there can only be one chiral fermion, we can get an expression for $C_{k_{\parallel}}$ by imposing that:


\[
1 = \int_{0}^{\infty}\bigl|\Psi_{k_{\parallel}}^{\gamma}(\bm{r})\bigr|^{2}\,dr
=|C|^{2}
\sum_{\eta,\eta'}
\biggl(-\frac{1}{2\bigl(\kappa^{\gamma\eta}+\kappa^{\gamma\eta'\,*}\bigr)}\biggr)
\]

\[
C =
\begin{cases}
\displaystyle
\Biggl[\,
\sum_{\eta,\eta'}
\Bigl(-\tfrac{1}{2(\kappa^{-,\eta}+\kappa^{-,\eta'\,*})}\Bigr)
\Biggr]^{-\tfrac12},
& -k_{F}<k_{\parallel}<0,\\[1.25em]
\displaystyle
\Biggl[\,
\sum_{\eta,\eta'}
\Bigl(-\tfrac{1}{2(\kappa^{+,\eta}+\kappa^{+,\eta'\,*})}\Bigr)
\Biggr]^{-\tfrac12},
& 0<k_{\parallel}<k_{F}.
\end{cases}
\]

\begin{figure}[H]
\centering
\includegraphics[width=0.25\textwidth]{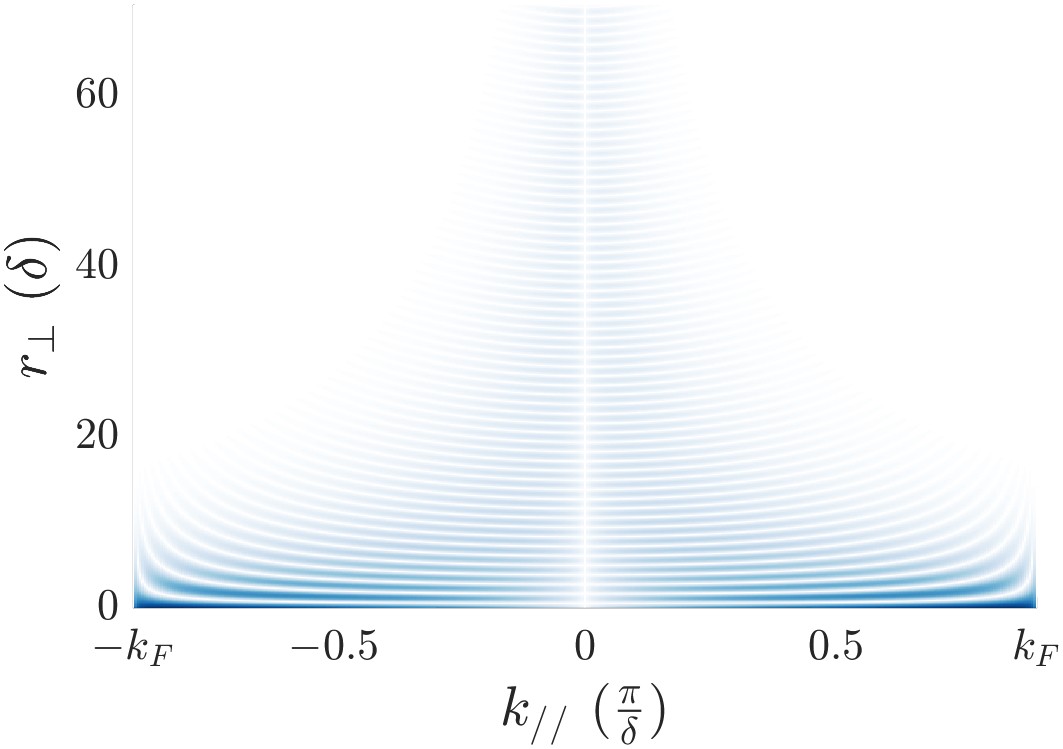}
\includegraphics[width=0.15\textwidth]{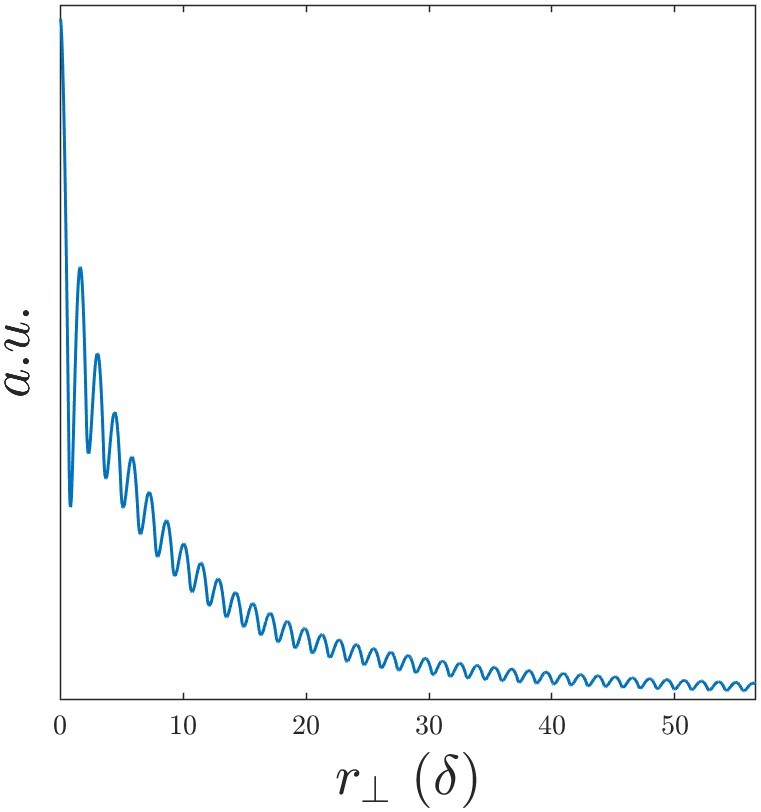}\\
\includegraphics[width=0.2\textwidth]{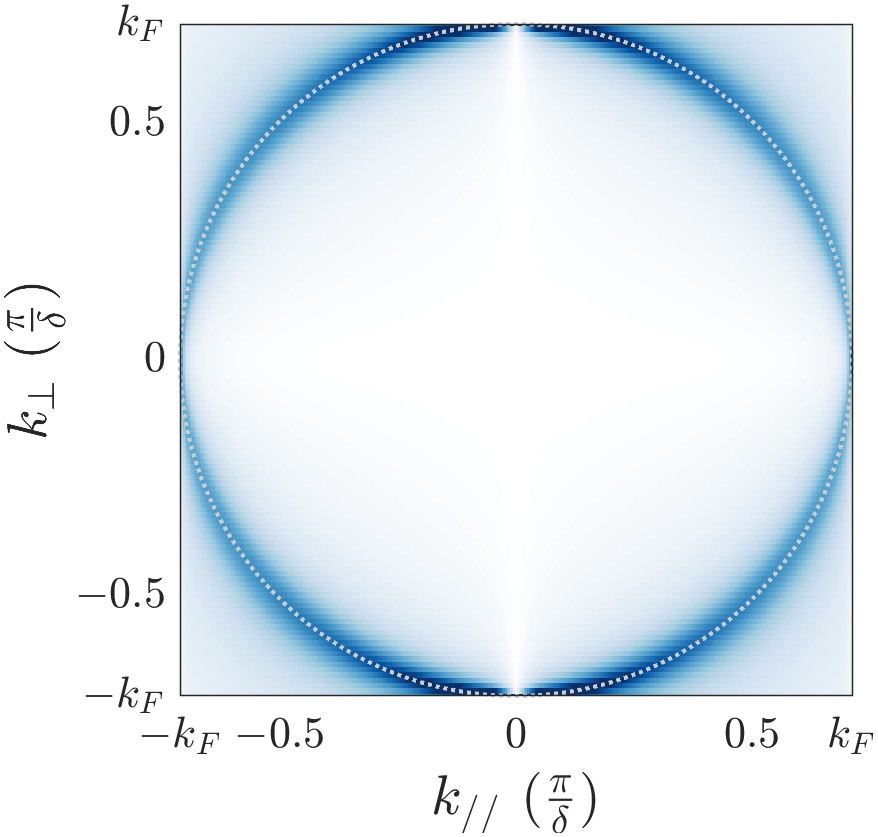}
\caption{\label{fig:Psi}(Top left) Absolute value of the Fourier transform of the wave function along $k_{\parallel}$, $\mathcal{F}_{k_{\parallel}}\left( \Psi_{k_{\parallel}}(\bm{r}) \right) = \Psi(k_{\parallel},r_{\perp})$ and its integration along $k_{\parallel}$ (top right). (Bottom) Absolute value of the full Fourier transform of the wave function $\mathcal{F}_{\bm{k}}\left( \Psi_{k_{\parallel}}(\bm{r}) \right) = \Psi(\bm{k})$. In all cases we used $|\Delta|/t=0.1$.}
\end{figure}

The wave function that we obtained is plotted in Figure \ref{fig:Psi}.

\subsection{\label{Supplementary:OOP_momentum_Dependence}Out-of-plane momentum dependence}

Even though we do not impose periodic boundary conditions in the direction perpendicular to the edge, in practice, because for most superconductors $\Delta \ll t$, to a first approximation, one can still obtain a relatively well defined momentum perpendicular to the edge. In fact, if we look at the Fourier transform of $e^{\kappa^{\gamma \eta} r_{\perp}}$:

\begin{align}
\mathcal{F}\left( e^{\kappa^{\gamma\eta} r_{\perp}}\right) = \frac{2\mathfrak{R} \left( \kappa^{\gamma\eta} \right)}{\mathfrak{R} \left( \kappa^{\gamma\eta} \right)^{2} + k_{\perp}^{2}} \ast \delta\left( k_{\perp}-\mathfrak{I} \left(\kappa^{\gamma\eta} \right) \right),
\end{align}

where to limit ourselves to real expressions we have also included the solutions for $r_{\perp}<0$ satisfying $\Psi_{k_{\parallel}}(r_{\parallel},r_{\perp} =\infty) = 0$, we see that we simply have a convolution of a Lorentzian $\mathcal{L}(k_{\perp};0,\mathfrak{R} \left( \kappa^{\gamma \eta} \right))$ and a Dirac delta function. Using the sifting property of the Dirac delta function and the fact that $\lim_{\gamma\to0} \mathcal{L}(x;x_{0},\gamma) = \delta\left( x-x_{0}\right)$, we get that:

\begin{align}
\lim_{\Delta/t\to0} \mathcal{F}\left( e^{\kappa^{\gamma \eta} r_{\perp}}\right) = \delta\left( k_{\perp} -\mathfrak{I} \left(\kappa^{\gamma \eta} \right)  \right),
\label{small_gap_limit}
\end{align}

which means that, as expected, for small gap-to-bandwidth ratios, the dependence of the spectral weight on $k_{\perp}$ becomes non-negligible. By equating Equation~\ref{eq:BdG_hamiltonian_epsilon} to zero, we can derive the expression of the perpendicular component of the Fermi momentum:

\begin{align}
k_{\perp}^{F} =  \frac{1}{t} \sqrt{t^2{\frac{t(4t+\mu)}{a^2} - k_{\parallel} }^2}.
\label{eq:k_perp_F}
\end{align}

We can then easily verify the equivalence:

\begin{align}
\lim_{\Delta/t\to0} \mathfrak{I} \left(\kappa^{\gamma \eta} \right) = k_{\perp}^{F}.
\label{Im_equivalence}
\end{align}

This means that in the limit of a small superconducting gap, the spectral weight of the edge state is transferred to the Fermi surface of the normal state. This is directly visible in Fig.~\ref{fig:Psi}, where because of the expansion to second order, the Fermi surface is a circle.

\subsection{\label{Supplementary:Scattering}Scattering sectors}

Consider a random onsite disorder acting on each site of the form expressed in the chiral basis:

\begin{align}
V_{q} = \sum_{\bm{r}} v(\bm{r}) \, \sigma_x \, e^{-i \bm{q} \cdot \bm{r}},
\end{align}

we can verify that it preserves the chiral symmetry of the Hamiltonian \ref{eq:BdG_hamiltonian} by computing its anticommutation relation with the chiral symmetry operator which in this basis is simply $\sigma_z$:

\begin{align}
\{ \sigma_z , \sigma_x \} = 0.
\end{align}

The chirality eigenstates are:
\[
\ket{+} = \begin{pmatrix}1 \\ 0\end{pmatrix}, \quad
\ket{-} = \begin{pmatrix}0 \\ 1\end{pmatrix}.
\]

So that the matrix elements of $\sigma_x$ are:
\begin{align}
\braket{\gamma | \sigma_x | \gamma'} 
= 1 - \delta_{\gamma,\gamma'}, 
\qquad \gamma \in \{+,-\}.
\end{align}
Thus a random onsite potential that preserves the chiral symmetry of the Hamiltonian mediates scattering only between states of opposite chirality. Scattering within the same chirality sector is instead forbidden. 

\subsection{\label{Supplementary:LAUE}LAUE pattern}

\begin{figure}[H]
\centering
\includegraphics[width=1\textwidth]{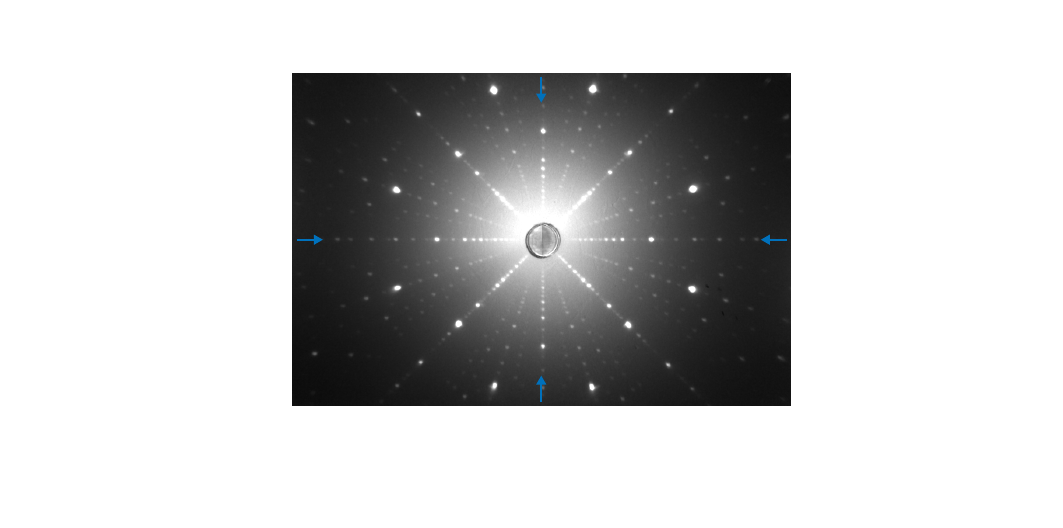}
\caption{Laue back-scattering diffraction pattern from a (001) surface used to orient the crystals along the [110] direction, indicated by the blue arrows.}
\end{figure}

\subsection{\label{Supplementary:SC_001}Superconducting gap on the (001) surface for a sample ($x = 0.19$) prepared under the same conditions}

  \begin{minipage}{0.43\textwidth}
    \centering
    \includegraphics[width=\linewidth]{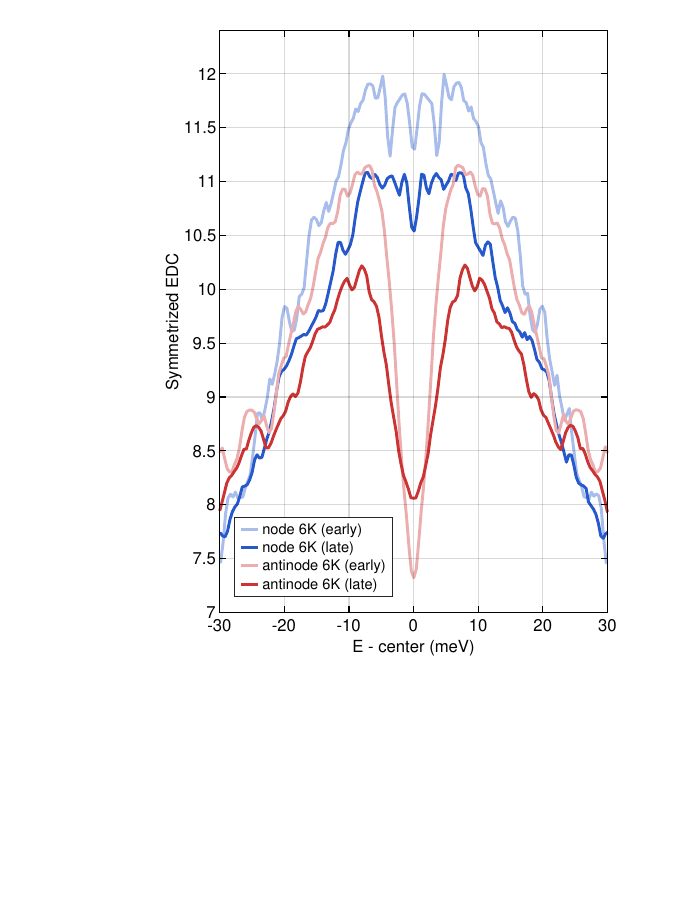}
  \end{minipage}\hfill
  \begin{minipage}{0.43\textwidth}
    \centering
    \includegraphics[width=\linewidth]{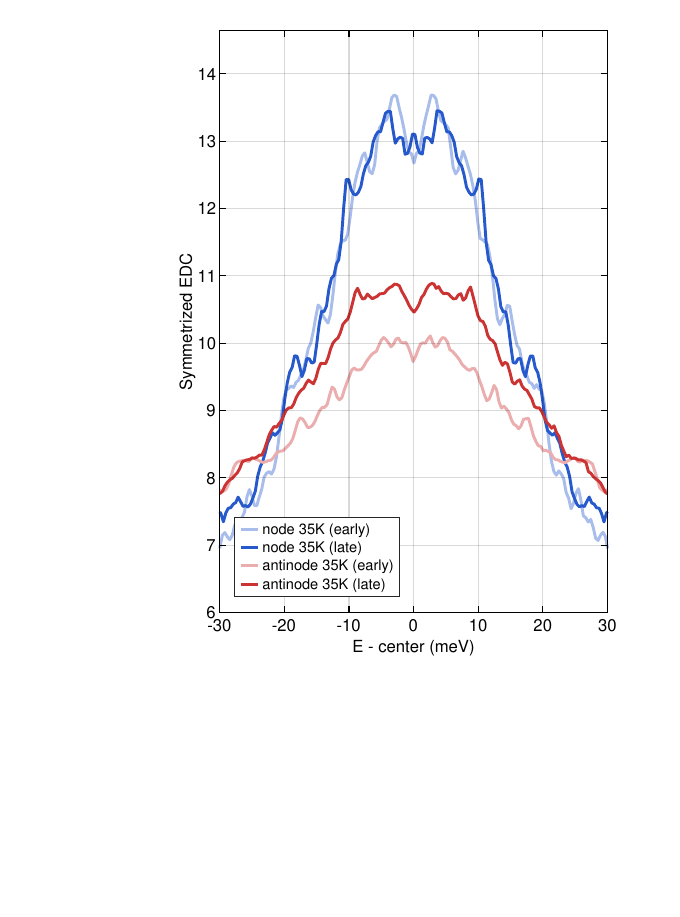}
    \end{minipage}
    \vspace{1em} 

  \begin{minipage}{0.43\textwidth}
    \centering
    \includegraphics[width=\linewidth]{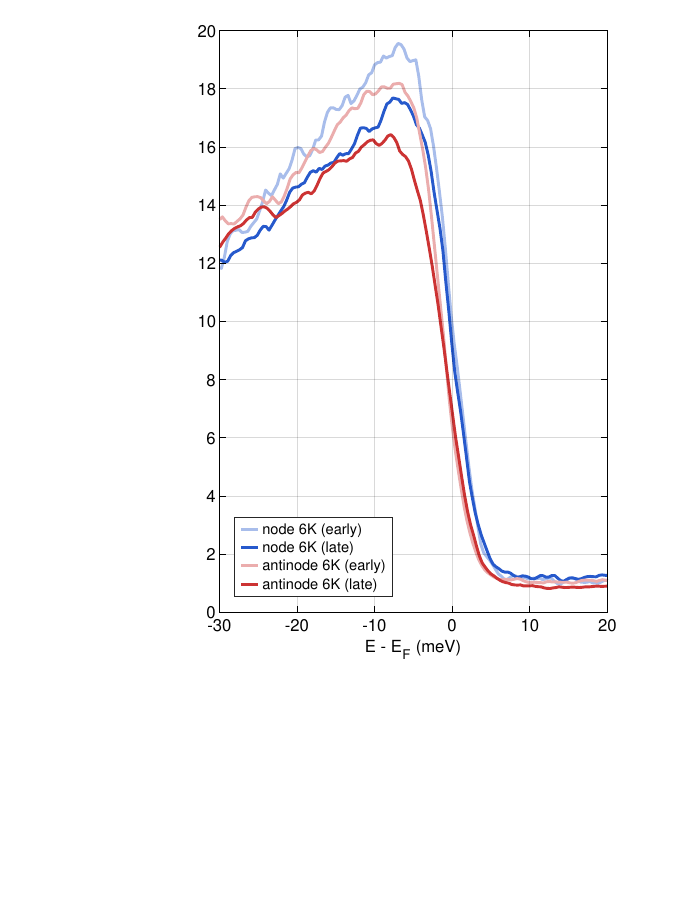}
  \end{minipage}\hfill
  \begin{minipage}{0.43\textwidth}
    \centering
    \includegraphics[width=\linewidth]{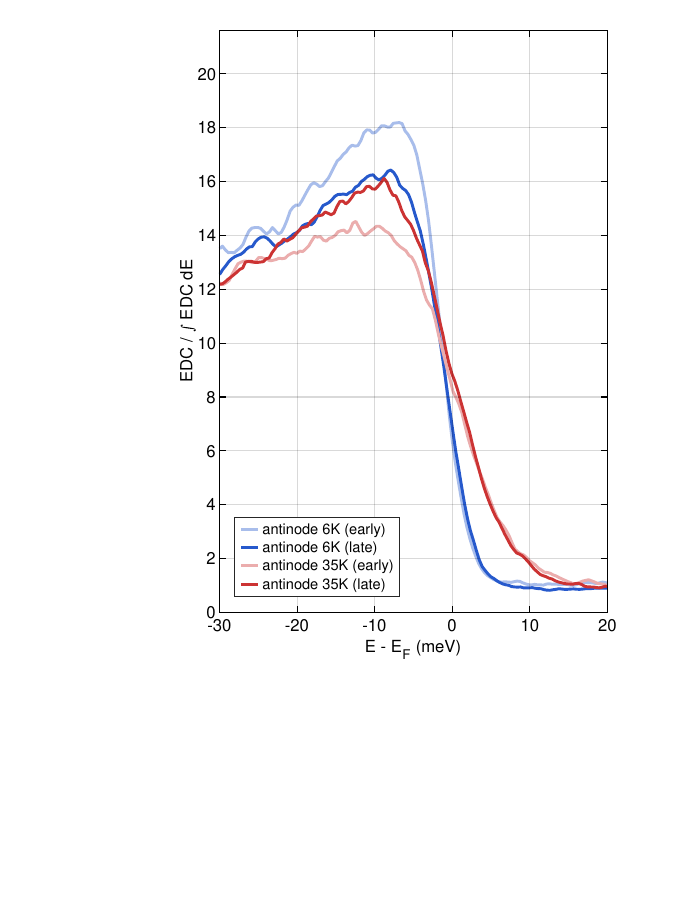}
  \end{minipage}

All spectra were first corrected for the instrumental Fermi-edge shape using a gold reference measured under identical conditions. Energy distribution curves (EDCs) at the nodal points were extracted in the second Brillouin zone, while both nodal and antinodal spectra were measured with LH polarization at a photon energy of 65 eV with a combined resolution of approximately $4$-$5$ meV. The nodal spectra were used as an internal reference in order to minimize differences in space-charge and conventional charging effects between the gold reference and the sample. Accordingly, the leading-edge midpoint (LEM) of the nodal EDCs at 6 K (and analogously at 35 K) were mutually aligned, and the corresponding shifts were applied to the antinodal spectra. The measurements were performed over two temperature cycles (“early” and “late”), and gold reference spectra were acquired throughout the experiment; fitting the Fermi edge with a Fermi–Dirac distribution convoluted with a Gaussian resolution function yielded a standard error of 0.125 meV, allowing the Fermi level to be treated as effectively constant. Additional flux-dependence measurements on the gold reference and on the sample at the nodes and antinodes were used to extrapolate the LEM and EF to zero flux, correcting for residual space-charge effects by 0.49 meV and 0.61 meV, respectively. The corrected spectra were subsequently symmetrized, yielding a gap of approximately 7 meV from the quasiparticle peak positions in the symmetrized EDCs. While this value is somewhat smaller than some reported in the literature obtained by fitting the symmetrized EDCs with a self-energy–based spectral function including resolution broadening~\cite{Zhong_2022}, it is comparable to others~\cite{Yoshida_2016}, where the gap was extracted by Fermi-function and normal-state division rather than symmetrization. Given the strong sensitivity of both the symmetrization and the division to small $E_{F}$ shifts, we consider our results to be consistent with previous reports, with the additional observation of a small LEM shift of a couple of meV.

\subsection{\label{Supplementary:Luttinger}Estimation of the doping level}

The Fermi momentum at the nodes is found to vary weakly with doping~\cite{Yoshida_2007}, meaning that an estimate of the surface doping from this single fitted value is likely inaccurate. However, the apparent distortion of the obtained Fermi surface, likely due to a combination of typical limitations of photon energy maps such as kz broadening, deviations from the free-electron approximation for the final state, small changes in space-charge effects with photon energy, as well as adjustments in the beamline optics, prevents us from obtaining an accurate estimate of the effective doping at the measured surface.

Nonetheless, we still obtained some indicative value of the doping by fitting a Fermi surface obtained from a tight-binding model to a set of points of the Fermi surface obtained from fitting gaussian peaks along various data slices. 
Because in a body-centered lattice neighboring unit cells are staggered, the photon energy dependence is obtaining cross-sections of the three-dimensional Fermi surface both at the $\Gamma$ and $Z$ planes. Thus, the hole doping, defined as the deviation from half filling, was obtained by computing the average ratio between the area contained by the Fermi pockets and the area of the Brillouin zone at the corresponding $\Gamma$ and $Z$ planes. This is particularly important for the Lanthanum family of cuprates where there is a non-negligible dispersion along the $c-axis$ direction~\cite{Fang_2022}.
The base model was the same as defined in Ref. ~\cite{Markiewicz_2005}, but to accommodate for the small distortions of our Fermi surface we added an additional parameter to allow for small deviations from the four-fold symmetry. Finally, in the fitting procedure we did not include points at different binding energies but only used the values of $k_F$.

This procedure gave us an estimate of the doping at the surface of $p = 0.2271$. Even though the fitted curve does not perfectly matches the data, this gives us a rough indication that the doping at the surface is not substantially different from the one in the bulk. Varying the constraints on the fitted parameters and with different initial guesses, we only obtained doping estimates in the slightly overdoped regime, up to a maximum of about $p = 0.3$ for the worst fits.

\begin{figure}[H]
\centering
\includegraphics[width=0.8\textwidth]{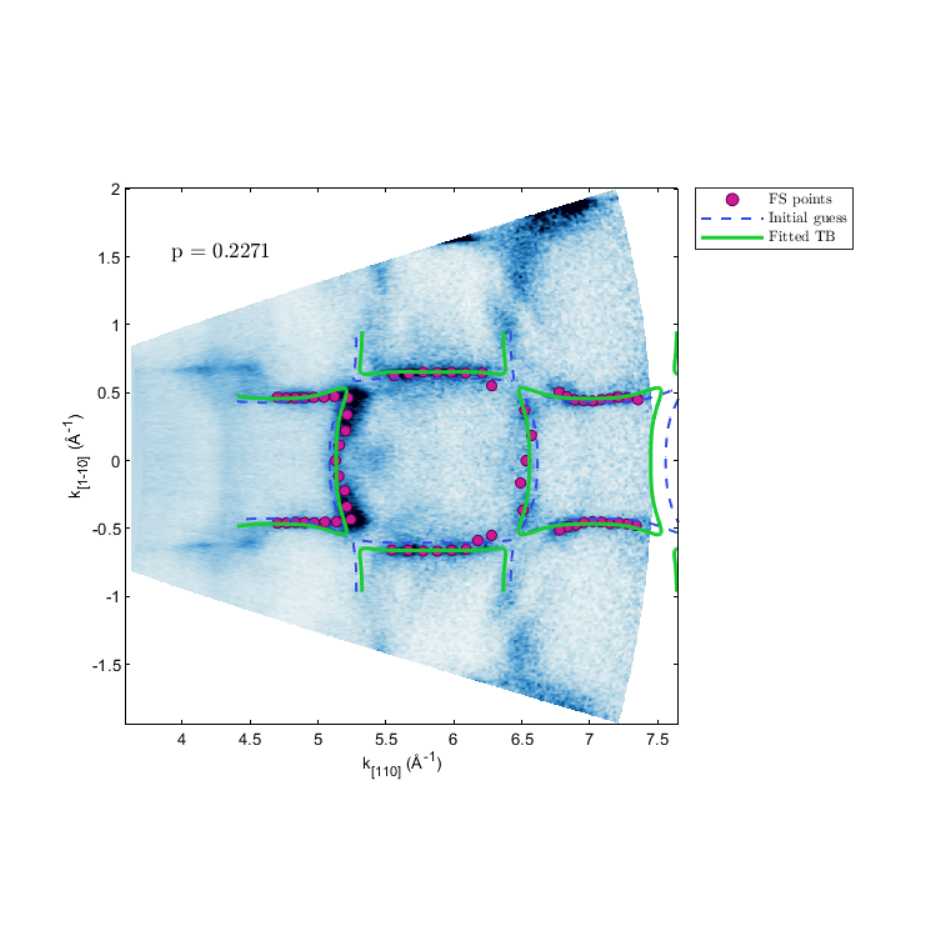}
\caption{\textbf{Fermi surface fitting} Photon energy map with the points of the Fermi surface used for the fitting (red circles), the initial guess (blue dashed line) and the converged result (green line).}
\end{figure}

\subsection{\label{Supplementary:EDC_boundary}EDC integration ranges}

\begin{figure}[H]
  \centering

  \begin{minipage}{0.3\textwidth}
    \centering
    \includegraphics[width=\linewidth]{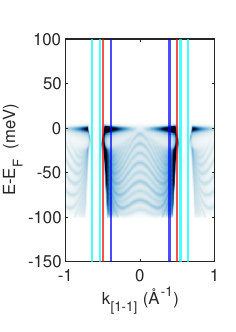}
  \end{minipage}\hfill
  \begin{minipage}{0.3\textwidth}
    \centering
    \includegraphics[width=\linewidth]{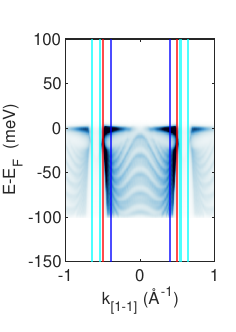}
  \end{minipage}\hfill
  \begin{minipage}{0.3\textwidth}
    \centering
    \includegraphics[width=\linewidth]{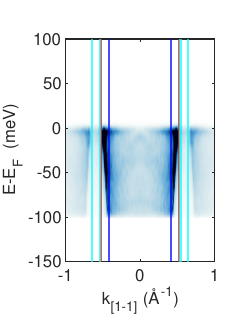}
  \end{minipage}

  \vspace{1em} 

  \begin{minipage}{0.3\textwidth}
    \centering
    \includegraphics[width=\linewidth]{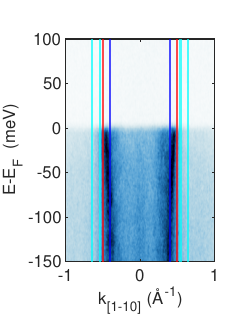}
  \end{minipage}\hfill
  \begin{minipage}{0.3\textwidth}
    \centering
    \includegraphics[width=\linewidth]{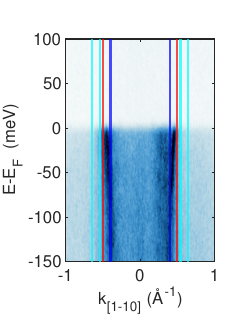}
  \end{minipage}\hfill
  \begin{minipage}{0.3\textwidth}
    \centering
    \includegraphics[width=\linewidth]{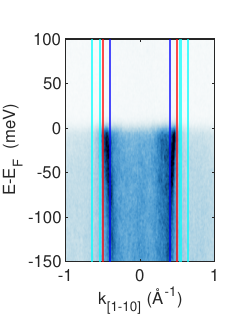}
  \end{minipage}

  \caption{Illustration of the integration ranges for the various EDCs. The region between the cyan and red lines corresponds to the trivial sector ($\nu = 0$), the region between the red and blue line corresponds to the antinode while the region between the blue bands corresponds to the topological sector ($\nu = \pm 1$).}
  \label{fig:EDC_boundary}
\end{figure}

\subsection{\label{Supplementary:Additional_EDCs}Additional EDCs}

\begin{figure}[H]
  \centering

  \begin{minipage}{0.32\textwidth}
    \centering
    \includegraphics[width=\linewidth]{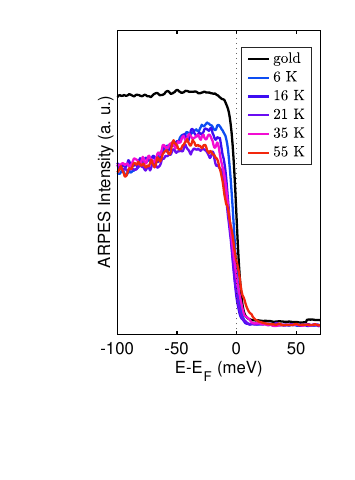}
  \end{minipage}\hfill
  \begin{minipage}{0.32\textwidth}
    \centering
    \includegraphics[width=\linewidth]{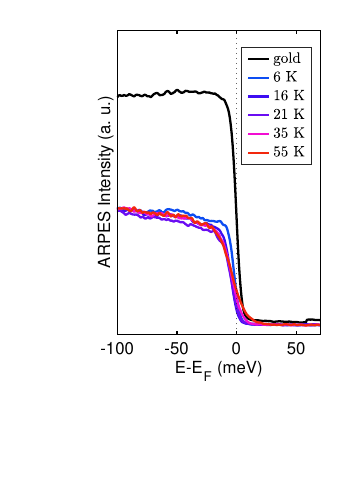}
  \end{minipage}\hfill
  \begin{minipage}{0.32\textwidth}
    \centering
    \includegraphics[width=\linewidth]{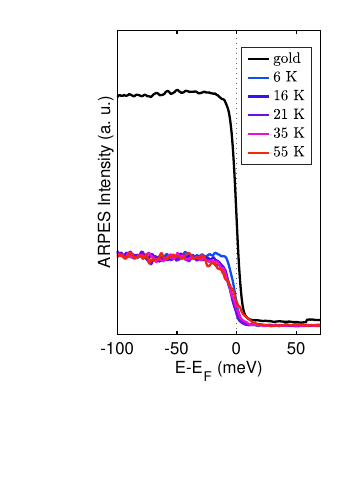}
  \end{minipage}

  \caption{EDCs obtained the same way as in Fig.~\ref{figure_3}e,g for a different sample.}
  \label{fig:Additional_EDCs}
\end{figure}

\subsection{\label{Supplementary:gap_001_GE+BA}Spatially resolved density of states}

\begin{figure}[H]
  \centering

  \begin{minipage}{0.4\textwidth}
    \centering
    \includegraphics[width=\linewidth]{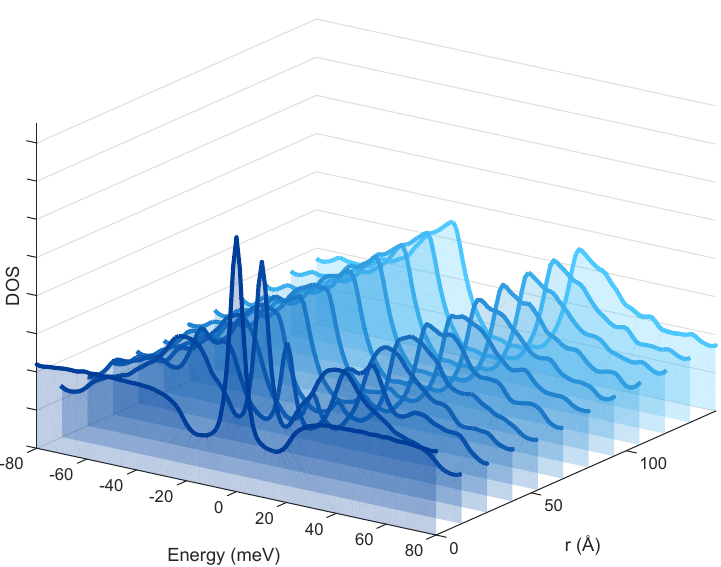}
  \end{minipage}\hfill
  \begin{minipage}{0.4\textwidth}
    \centering
    \includegraphics[width=\linewidth]{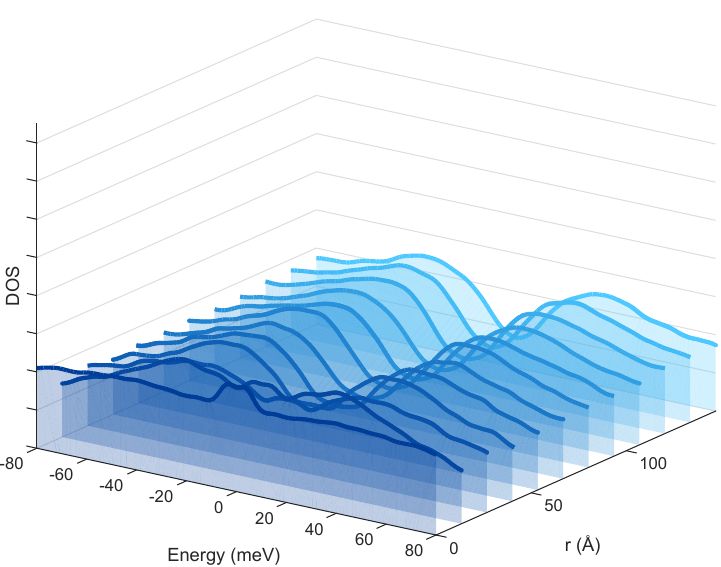}
  \end{minipage}

  \caption{Spatially-resolved density of states for the clean (left) and the GE+BA disordered (right) sample, showing that one can have a bulk gap even for disorder magnitudes that lift the degeneracy of the surface states.}
  \label{fig:real_space_DOS}
\end{figure}

\subsection{\label{Supplementary:edge_w_and_wo_gap_suppression}Effect of the gap suppression on the edge states}

\begin{figure}[H]
  \centering

  \begin{minipage}{0.4\textwidth}
    \centering
    \includegraphics[width=\linewidth]{Figures/DOS_layers_3D_Clean.pdf}
  \end{minipage}\hfill
  \begin{minipage}{0.4\textwidth}
    \centering
    \includegraphics[width=\linewidth]{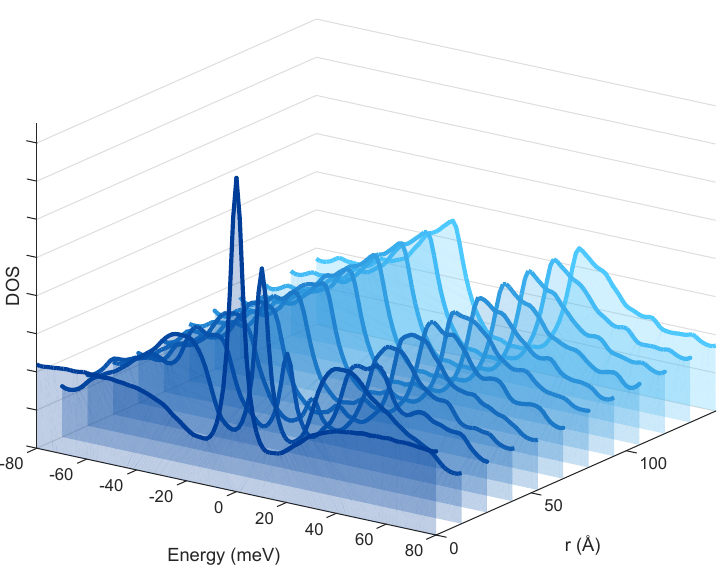}
  \end{minipage}

  \caption{Spatially-resolved density of states for a clean system with gap suppression at the edge (left) and without self consistency of the gap equation, i.e. the gap at the edge is equal to the gap in the bulk (right). The effect of the gap suppression on the spectral function is to have an edge state slightly more delocalized into the bulk as well as narrowing the "gap-like" dip at the very edge.}
  \label{fig:real_space_DOS_wo_self_C}
\end{figure}

\newpage

\end{document}